\documentclass[12pt]{elsarticle}

\usepackage{etoolbox}

\usepackage{color}
\usepackage{amsmath}
\usepackage[dvipsnames]{xcolor}
\usepackage{soul}
\usepackage[caption=false]{subfig}
\sethlcolor{orange}

\usepackage{graphicx}
\usepackage{amssymb}

\usepackage{lineno}

\journal{Engineering Analysis with Boundary Elements}

\begin{document}

\makeatletter
\patchcmd{\ps@pprintTitle}
  {Preprint submitted to}
  {Published in}
  {}{}
\makeatother

\begin{frontmatter}


\title{Eliminating the fictitious frequency problem in BEM solutions of the external Helmholtz equation} 





\author[1]{Evert Klaseboer}
\address[1]{Institute of High Performance Computing, 1 Fusionopolis Way, Singapore 138632, Singapore}
\author[2]{Florian D. E. Charlet}
\author[2]{Boo-Cheong Khoo}
\address[2]{Department of Mechanical Engineering, National University of Singapore 117575, Singapore}

\author[3,4]{\\Qiang Sun\corref{correfSQ}}
\address[3] {Department of Chemical Engineering, The University of Melbourne, Parkville 3010, VIC, Australia}
\address[4] {ARC Centre of Excellence for Nanoscale BioPhotonics (CNBP), School of Science, RMIT University, Melbourne, VIC 3001, Australia}
\cortext[correfSQ]{qiang.sun@rmit.edu.au}

\author[5,6]{Derek Y. C. Chan}
\address[4]{School of Mathematics and Statistics, The University of Melbourne, Parkville 3010, VIC, Australia}
\address[5]{Department of Mathematics, Swinburne University of Technology, Hawthorn VIC 3121 Australia}

\begin{abstract}

The problem of the fictitious frequency spectrum resulting from numerical implementations of the boundary element method for the exterior Helmholtz problem is revisited. When the ordinary 3D free space Green's function is replaced by a modified Green's function, it is shown that these fictitious frequencies do not necessarily have to correspond to the internal resonance frequency of the object. Together with a recently developed fully desingularized boundary element method that confers superior numerical accuracy, a simple and practical way is proposed for detecting and avoiding these fictitious solutions.  The concepts are illustrated with examples of a scattering wave on a rigid sphere.
\end{abstract}

\begin{keyword}
Internal resonance \sep desingularized boundary element method \sep Frequency shift \sep Modified Green's function


\end{keyword}

\end{frontmatter}


\section{Introduction}
\label{sec:Introduction}

Recent studies of the boundary integral formulation of problems in time domain acoustic scattering \cite{KlaseboerJASA2017}, wave propagation in periodic structures~\cite{Fu2018}, dynamic elasticity using the Helmholtz decomposition method \cite{KlaseboerJE2018} and direct field-only formulation of computational electromagnetics \cite{KlaseboerIEEETAP2017, SunPRB2017, KlaseboerAO2017}, all rely on finding accurate and efficient methods of solving the scalar Helmholtz equation. In this regard, it is timely to re-visit the boundary integral method of solving the Helmholtz equation.

It is well-known that the solution of the Helmholtz equation for external problems obtained by the boundary integral method, BIM, (or its numerical counterpart: boundary element method BEM) can become non-unique at certain frequencies (similar problems can occur for multiply connected domains \cite{ChenRoyalSoc2001}). At these so called fictitious \cite{ChenEABE2009} or irregular \cite{ZouEABE2018} frequencies, the non-physical solutions that arise are said to correspond to the internal resonance frequencies of the scatterer. Although there are established methods, most notably due to Schenck \cite{schenck1968} and to Burton and Miller \cite{BurtonMiller1971} that have been developed to eliminate such fictitious solutions, these methods require numerical tools beyond the BIM. For instance, the solution of Schenck requires additional numerical algorithms such as least squares minimization and that of Burton and Miller leads to hypersingular integral equations \cite{Li2010, Fu2014, Zheng2015, Feng2015}. Here we show that these fictitious solutions, when they do occur, and their corresponding frequencies in the BIM context depend not only on the shape of the object but also on the choice of Green's function so that these frequencies do not necessarily occur at the corresponding internal resonance frequencies of the object. This observation together with the fact that recently developed desingularized BIM can give sufficiently high precision that the solution is unaffected by such fictitious solutions until the frequency is within about 1 part in $10^4$ of a fictitious value. We shall demonstrate how this can be exploited to detect the presence of a fictitious solution. Furthermore, the fictitious frequency spectrum can be changed by using different Green's functions in the BIM. Taken together, these developments provide a practical way to detect and eliminate the effects of the fictitious solution without additional numerical effort or adjustable parameters beyond the toolkit of the BIM. 

The introduction of a modified Green's function also poses a number of interesting but unanswered questions that can provide stimulus for further theoretical development.

To provide physical context to our discussion on how the fictitious solution arises in the solution of the Helmholtz wave equation using the boundary integral method, we consider the example of the scattering of an incident acoustic wave by an object with boundary $S$ in an infinite medium. In the external domain, assumed to be homogeneous, scattered acoustic oscillations are described by the Helmholtz scalar wave equation in the frequency domain:
\begin{equation} \label{eq:Helmholtzeq}
    \nabla^2 \phi + k^2 \phi =0,
\end{equation}
where $k = \omega/c$ is the wave number, $\omega$ the angular frequency and $c$ the speed of sound. The (complex) acoustic potential, $\phi$, is related to the scattered velocity: $\boldsymbol u = \nabla \phi$. Since Eq.~\ref{eq:Helmholtzeq} is elliptic, the Green's function formalism can be used to express the solution as that of a boundary integral equation \cite{Kinsler, Kirkup}
\begin{equation} \label{eq:BEM1}
c(\boldsymbol x_0)\phi(\boldsymbol x_0)+\int_S \phi(\boldsymbol x)
\frac{\partial G(\boldsymbol x ,\boldsymbol x_0 | k)}
{\partial n} \text{ d}S(\boldsymbol x ) = \int_S \frac{\partial \phi(\boldsymbol x)}{\partial n} G (\boldsymbol x, \boldsymbol x_0 | k)\text{ d}S(\boldsymbol x ),
\end{equation}
where 
\begin{equation} \label{eq:Green1}
    G(\boldsymbol x ,\boldsymbol x_0 | k)=
    \frac{e^{ikr}}{r}
\end{equation} 
is the 3D Green's function with $ r=\|\boldsymbol x -\boldsymbol x_0 \|$  and $\partial / \partial n \equiv \boldsymbol{n} \cdot \nabla$ is the normal derivative where the normal vector $\boldsymbol n$ points out of the domain, and thus into the object. The position vector $\boldsymbol x$ in Eq.~\ref{eq:Helmholtzeq} is located on the boundary $S$. If the observation point $\boldsymbol x_0$ is located outside the object (i.e. \emph{within} the solution domain), the solid angle $c = 4\pi$, if $\boldsymbol x_0$ is located inside the object (i.e. \emph{outside} the solution domain), $c = 0$, and if $\boldsymbol x_0$ is located on the surface, $S$, of the object and that point on $S$ has a continuous tangent plane, then and only then $c = 2\pi$, otherwise the value of the solid angle $c$ is determined by the local surface geometry at $\boldsymbol x_0$.

The advantages of using Eq.~\ref{eq:BEM1} over other methods such as using finite difference in the 3D domain are the obvious reduction in the spatial dimension by one and that it is relatively easy to accommodate complicated shapes without deploying multi-scale 3D grids. Also the Sommerfeld radiation condition at infinity \cite{Schot1992} is automatically satisfied by Eq.~\ref{eq:BEM1}.

For the simple example of the scattering of an incoming plane wave specified by $\phi^{\text{inc}}=\Phi_0 e^{i\boldsymbol{k}\cdot \boldsymbol x}$ (with $\Phi_0$ a constant and $\| \boldsymbol k\|=k$) by a rigid object, the velocity potential, $\phi$ of the scattered wave can be found by solving Eq.~\ref{eq:Helmholtzeq}. The condition of zero normal velocity on the surface is equivalent to the boundary condition on $S$: $\partial \phi /\partial n = -\partial \phi^{\text{inc}}/ \partial n$. In this case, the right hand side of Eq.~\ref{eq:BEM1} is known so this equation can be solved for the velocity potential, $\phi(\boldsymbol x_0)$, with $\boldsymbol x_0$ on the surface. 

We now demonstrate using this example of a Neumann problem where $\partial \phi/\partial n$ is given on the surface $S$, that there exists certain values of $k = k_f$, at which the solution $\phi$ of Eq.~\ref{eq:BEM1} is no longer unique. This occurs at those frequencies $k_f$ whereby a non-trivial function $f$ can exist to satisfy the following homogeneous equation:
\begin{equation} \label{eq:BEM2}
c(\boldsymbol{x}_0)f(\boldsymbol x_0 | k_f)+\int_S f(\boldsymbol x | k_f)
\frac{\partial G(\boldsymbol x ,\boldsymbol x_0 | k_f)}
{\partial n} \text{ d}S = 0.
\end{equation}
Consequently Eq.~\ref{eq:BEM1} will admit a solution of the form $\phi + b f$ on the surface $S$, where $b$ is an arbitrary constant and $f$, the fictitious solution, also satisfies the integral equation with zero normal derivative on $S$. Thus the fictitious frequency, $k_f$ and the corresponding fictitious solution, $f(\boldsymbol x | k_f)$ are the eigenvalue and eigenfunction of Eq.~\ref{eq:BEM2}, respectively. The existence of fictitious frequencies in boundary integral methods for Helmholtz equations was already identified by Helmholtz in 1860 \cite{Helmholtz1860}, who said on page 24 (see also page 29 of his book \cite{Helmholtz}), while discussing the integral equation, Eq.~\ref{eq:BEM1}:
\begin{quote}...aber f\"{u}r eine unendlich grosse Zahl von bestimmten Werthen von $k$ f\"{u}r eine jede gegebene geschlossene Oberfl\"{a}che Ausnahmen erleidet. Es sind dies n\"{a}mlich diejenigen Werthe von $k$, die den eigenen T\"{o}nen der eingeschlossenen Luftmasse entsprichen.
\end{quote}
This text was more or less translated directly by Rayleigh \cite{bookRayleigh} in his book:
\begin{quote}For a given space $S$ there is .... a series of determinate values of $k$, corresponding to the periods of the possible modes of simple harmonic vibration which may take place within a closed rigid envelope having the form of $S$. With any of these values of $k$, it is obvious that $\phi$ cannot be determined by its normal variation over $S$, and the fact that it satisfies throughout $S$ the equation $\nabla^2 \phi + k^2 \phi =0$.
\end{quote}

Note that the internal resonance problem corresponds to a problem with $\phi=0$ on the surface, $S$ and $g \equiv \partial \phi /\partial n \neq 0$ in Eq.~\ref{eq:BEM1}, is given by
\begin{equation} \label{eq:IntRes1}
\int_S g(\boldsymbol x | k_f)
 G(\boldsymbol x ,\boldsymbol x_0 | k_f) \text{ d}S = 0,
\end{equation}
which is different from Eq.~\ref{eq:BEM2}. It is not immediately obvious that Eqs.~\ref{eq:BEM2} and \ref{eq:IntRes1} will produce the same fictitious spectrum and in fact, as we shall see later in Section~\ref{sec:ModifiedGreen}, this is not always the case.

In theory, the fictitious solution only appears if $k$ is \emph{exactly} equal to $k_f$ so that it is not an issue in analytic work nor if computations have infinite numerical precision. With the advent of numerical techniques in the late 1960's and early 1970's, the boundary integral equation was transformed into the boundary element method (BEM). The issue of fictitious frequencies now resurfaced once more in the numerical implementations. In the conventional implementation of the BEM \cite{Kirkup}, the surface $S$ is represented by a mesh of planar area elements and the unknown value of $\phi(\boldsymbol x)$ on the surface is assumed to be a constant within each planar element and only varies from element to element. The surface integral is thus converted to a linear system in which the values of $\phi$ at different area elements are unknowns to be solved. The practicality of discretization where the representation of the surface $S$ by a finite number of planer elements and round off errors in numerical computation mean that effects of the fictitious solution begin to be important, not only when $k=k_f$, but even when the value of $k$ is near $k_f$. For instance, in a conventional implementation of the BEM, the apparent location of the fictitious frequency, $k_f$ can be in error because of the approximation involved in representing the actual surface by a set of planar elements. Thus the mean relative error can exceed 100\% when $k$ is within 1-2\% of the actual fictitious frequency (see Fig.\ref{fig:FigBRIEFvsCBIM} for examples of a sphere with radius $R$ at $kR\approx \pi$ and $kR\approx 2\pi$). Since the values of $k_f$ are not known \emph{a priori} for general boundary shapes, $S$, the accuracy of any BEM solution of the Helmholtz equation can become problematic. 

Two popular methods to deal with this issue that are still in use today are due to Schenck \cite{schenck1968} and to Burton and Miller \cite{BurtonMiller1971}. Schenck introduced the CHIEF method whereby the BEM solution is evaluated at additional internal points inside the scatterer with the requirement that such values must vanish. This results in an over-determined matrix system that requires a least square solution entailing considerable additional computational time, especially for larger systems. However, the CHIEF method does not stipulate how many CHIEF points should be used and where they should be placed. The Burton and Miller \cite{BurtonMiller1971} method involves taking the normal derivative of Eq.~\ref{eq:BEM1}, multiplying it by an appropriate complex number and then adding it to the original equation. It is claimed that Eq.~\ref{eq:BEM1} and its normal derivative have different resonance spectra and this therefore solves the fictitious frequency problem. Due to the use of the normal derivative of Eq.~\ref{eq:BEM1}, the Burton and Miller method involves having to deal with strongly singular kernels. This approach therefore has the disadvantage that it requires special quadrature rules for higher order elements \cite{harris1992}.

The issue of fictitious solutions is revisited in this article. Clearly, if a numerical implementation of the BEM is not sensitive to the fact that $k$ may be close to a fictitious value $k_f$, then the effects of a fictitious solution will be minimized. Furthermore, the spectrum of fictitious frequencies does not only depend on the shape of the object, but also on the choice of the Green's function. As the classical free space Green's function or fundamental solution of Eq.~\ref{eq:Green1} is not the only choice that can be used, it can be replaced by other fundamental solutions, as long as they are analytic in the external domain and they satisfy the Sommerfeld radiation condition \cite{Ursell1973}. Thus using a different Green's function will shift the spectrum of fictitious frequencies relative to a given $k$ value. Although the theoretical framework of modified Green's functions has been discussed extensively in the literature \cite{Ursell1973,KleinmanRoach1982, Jones1973, Ursell1978, Martin2002},  only very little attention appears to have been paid to the actual implementation, for example, the cases of Neumann boundary condition and of Dirichlet boundary condition were considered in two nearly identical papers~\cite{Lin2002, Lin2004}. In this article we address this issue.

The development of our suggestion to eliminate the fictitious frequency problem in BEM solutions of the external Helmholtz equation is organized as follows. In Section~\ref{sec:NEBEMsolution}, we outline how a desingularized implementation of the BEM that is not affected by a fictitious solution unless $k$ is very close to a fictitious value $k_f$, can be used to decide if an BEM solution has been adversely affected by the presence of a fictitious component. This framework also enables us to implement higher order elements with ease. In Section~\ref{sec:spuriousorigins} the spectrum of fictitious frequencies and corresponding solutions are studied as the solution of an homogeneous integral equation. In Section~\ref{sec:ModifiedGreen}, a modified Green's function is introduced to show how it can be used to change the spectrum of fictitious frequencies. Thus by employing the desingularized BEM, it is sometimes easy to determine by comparing the solutions obtained from using the conventional Green's function in Eq.~\ref{eq:Green1}, and from a modified Green's function whether the solutions have been adversely affected by the presence of a solution associated with a fictitious frequency. Some discussion and the conclusion follow in Sections~\ref{sec:Discussion} and \ref{sec:Conclusions}, respectively.

\section{Minimize the proximity effects to a fictitious frequency} \label{sec:NEBEMsolution}

As noted earlier, discretization and round off errors can cause the spurious solution to become important when the wave number happens to be near a fictitious value. However, since the spectrum of fictitious frequencies is not generally known \emph{a priori}, the numerical accuracy of a solution obtained by the BEM becomes uncertain. Therefore, to ameliorate the fictitious frequency problem, it is valuable to have an accurate implementation of the BEM that will not produce a fictitious component to the solution unless the frequency $k$ is extremely close to an unknown fictitious frequency. This is provided by a recently developed fully desingularized boundary element formulation \cite{KlaseboerJFM2012, SunRoySoc2015}, a concept that was first introduced for the BEM solution of the Laplace equation by Klaseboer et al. \cite{KlaseboerEABE2009}. In this framework, the traditional singularities of the Green's function and its normal derivative in the BEM integrals are removed analytically from the start. 

High accuracy can be achieved in this approach firstly due to the fact that all elements (including the previously singular one) are treated in the same manner with the same Gaussian quadrature scheme. The second reason for the high accuracy lies in the fact that instead of using planar area elements in which the unknown functions are assumed to be constant within such elements, the unknowns are now function values at node points on the surface, and the surface is represented more accurately by quadratic area elements determined by these nodal points. In calculating integrals over the surface elements, variation of the function value within each element is also estimated by quadratic interpolation from the nodal values. The numerical implementation is straightforward, once the linear system is set up, the usual linear solvers can be used. The thus obtained framework is termed Boundary Regularized Integral Equation Formulation (or BRIEF in short \cite{SunRoySoc2015}).

Here is a brief description of the desingularized boundary element formulation, details of which are given in previous works \cite{KlaseboerJFM2012, SunRoySoc2015}. Assume we have a known analytic solution, $\Psi (\boldsymbol x)$, of the Helmholtz equation in Eq.~\ref{eq:Helmholtzeq} which then also satisfies Eq.~\ref{eq:BEM1} as: 
\begin{equation} \label{eq:BEM_Desing1}
\begin{aligned}
c\Psi(\boldsymbol x_0)+ \int_S \Psi(\boldsymbol x)
\frac{\partial  G(\boldsymbol x ,\boldsymbol{x}_{0} |k)}
{\partial n} \text{ d}S = \int_S \frac{\partial \Psi(\boldsymbol x)}{\partial n} G (\boldsymbol x, \boldsymbol{x}_{0} |k) \text{ d}S.
\end{aligned}
\end{equation}
Without loss of generality, we can demand that this solution further satisfies the following two point-wise conditions when $\boldsymbol{x}_{0}$ is on surface $S$:
\begin{equation} \label{eq:BEM_DesingBC1}
\lim_{\boldsymbol x \to \boldsymbol x_0} \Psi (\boldsymbol x) = \phi (\boldsymbol x_0)
\end{equation}
\begin{equation} \label{eq:BEM_DesingBC2}
\lim_{\boldsymbol x \to \boldsymbol x_0} \frac{\partial \Psi (\boldsymbol x)}{\partial n} = \frac{\partial \phi (\boldsymbol x_0)}{\partial n}
\end{equation}
A convenient but not the only possible choice is a combination of two standing waves, one with the node of the wave and the other with the antinode situated at $\boldsymbol x_0$, both aligned with $\boldsymbol n (\boldsymbol x_0)$ \cite{KlaseboerJFM2012} as:
\begin{equation}
\label{eq:BEM_Desing2}
\begin{aligned}
\Psi(\boldsymbol x) = \cos \big(k \boldsymbol n(\boldsymbol x_0) & \cdot [\boldsymbol x - \boldsymbol x_0]\big)\phi (\boldsymbol x_0)\\& + \frac{1}{k}\sin \big(k \boldsymbol n(\boldsymbol x_0) \cdot [\boldsymbol x - \boldsymbol x_0]\big)\frac{\partial\phi (\boldsymbol x_0)}{\partial n}.
\end{aligned}
\end{equation}
Substituting Eq.~\ref{eq:BEM_Desing2} in Eq.~\ref{eq:BEM_Desing1} and subtracting the result from Eq.~\ref{eq:BEM1} gives:
\begin{equation} \label{eq:BEM_Desing3}
\begin{aligned}
4\pi\phi(\boldsymbol x_0)+ \int_S \big[\phi(\boldsymbol x)-\Psi(\boldsymbol x)\big] &
\frac{\partial  G(\boldsymbol x ,\boldsymbol x_0 |k)}
{\partial n} \text{ d}S = \\ & \int_S \Big[\frac{\partial \phi(\boldsymbol x)}{\partial n}-\frac{\partial \Psi(\boldsymbol x)}{\partial n} \Big] G (\boldsymbol x, \boldsymbol x_0 |k) \text{ d}S.
\end{aligned}
\end{equation}
The conditions from Eqs.~\ref{eq:BEM_DesingBC1} and \ref{eq:BEM_DesingBC2} guarantee that the terms in $[...]$ on both sides of Eq.~\ref{eq:BEM_Desing3} cancel out the singularities of the Green's function and its derivative by noting that
\begin{equation}
\label{eq:BEM_Desing4}
\begin{aligned}
\frac{\partial\Psi(\boldsymbol x)}{\partial n} =\boldsymbol n \cdot\nabla \Psi= &-k \boldsymbol n (\boldsymbol x) \cdot \boldsymbol n (\boldsymbol x_0)\sin \big(k \boldsymbol n(\boldsymbol x_0) \cdot [\boldsymbol x - \boldsymbol x_0]\big)\phi (\boldsymbol x_0)\\& + \boldsymbol n (\boldsymbol x) \cdot \boldsymbol n (\boldsymbol x_0)\cos \big(k \boldsymbol n(\boldsymbol x_0) \cdot [\boldsymbol x - \boldsymbol x_0]\big)\frac{\partial\phi (\boldsymbol x_0)}{\partial n},
\end{aligned}
\end{equation}
and the fact that $\boldsymbol n (\boldsymbol x) \cdot \boldsymbol n (\boldsymbol x_0) \to 1$, when $\boldsymbol x$ approaches $\boldsymbol x_0$ for any smooth surface. Note that the solid angle in Eq.~\ref{eq:BEM_Desing3} has been eliminated, but a term with $4\pi \phi(\boldsymbol x_0)$ appears due to the contribution of the integral over a surface at infinity because of the particular choice of Eq.~\ref{eq:BEM_Desing2}. Also note from Eq.~\ref{eq:BEM_Desing2} that $\Psi$ is a different function for each node on the surface. It is noted that other desingularization methods based on entirely different concepts exist as well in the literature \cite{LiEABE2019}.

\begin{figure}[!t]
\centering
\subfloat[]{ \includegraphics[width=0.48\linewidth]{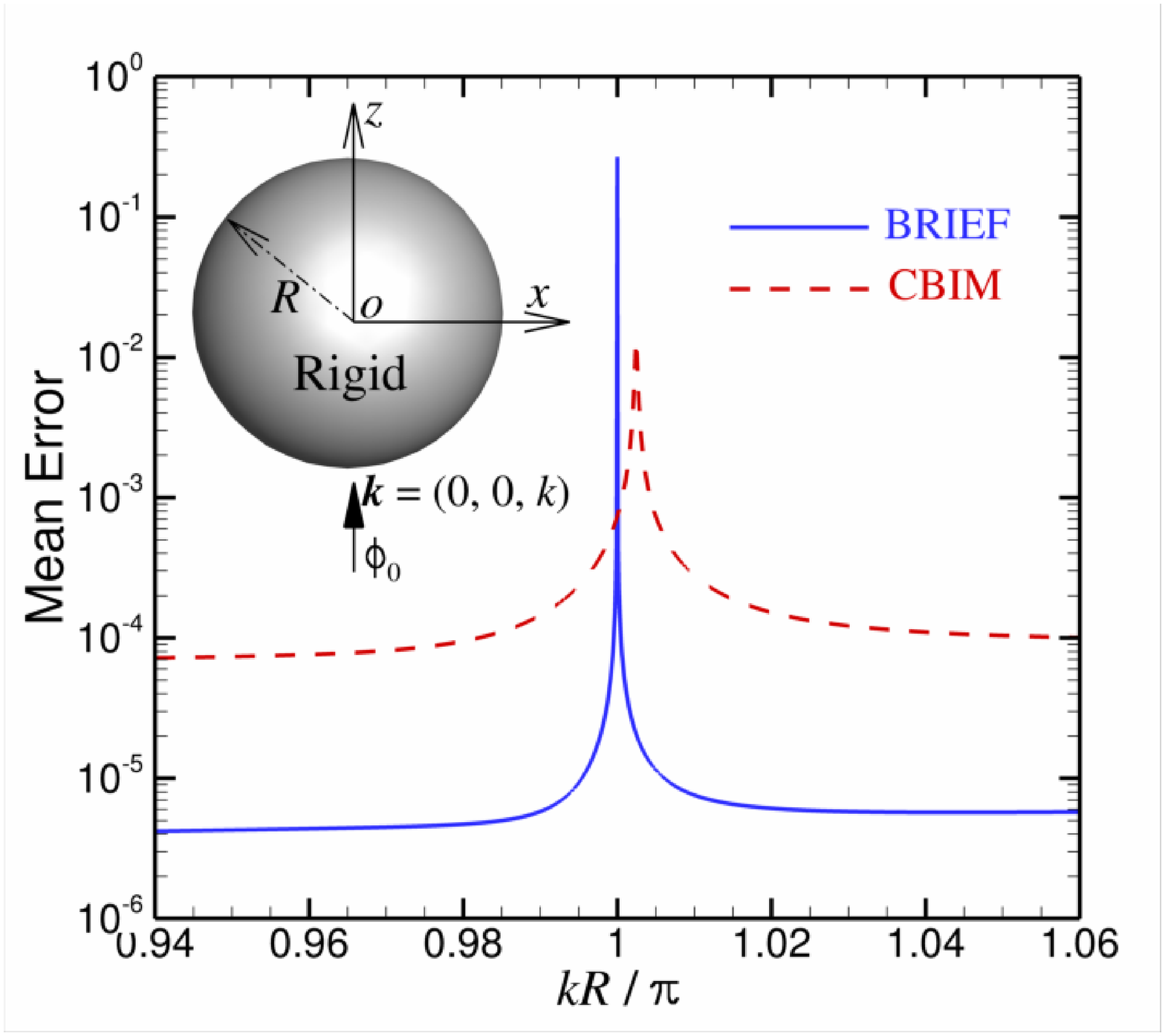} }
\subfloat[]{ \includegraphics[width=0.48\linewidth]{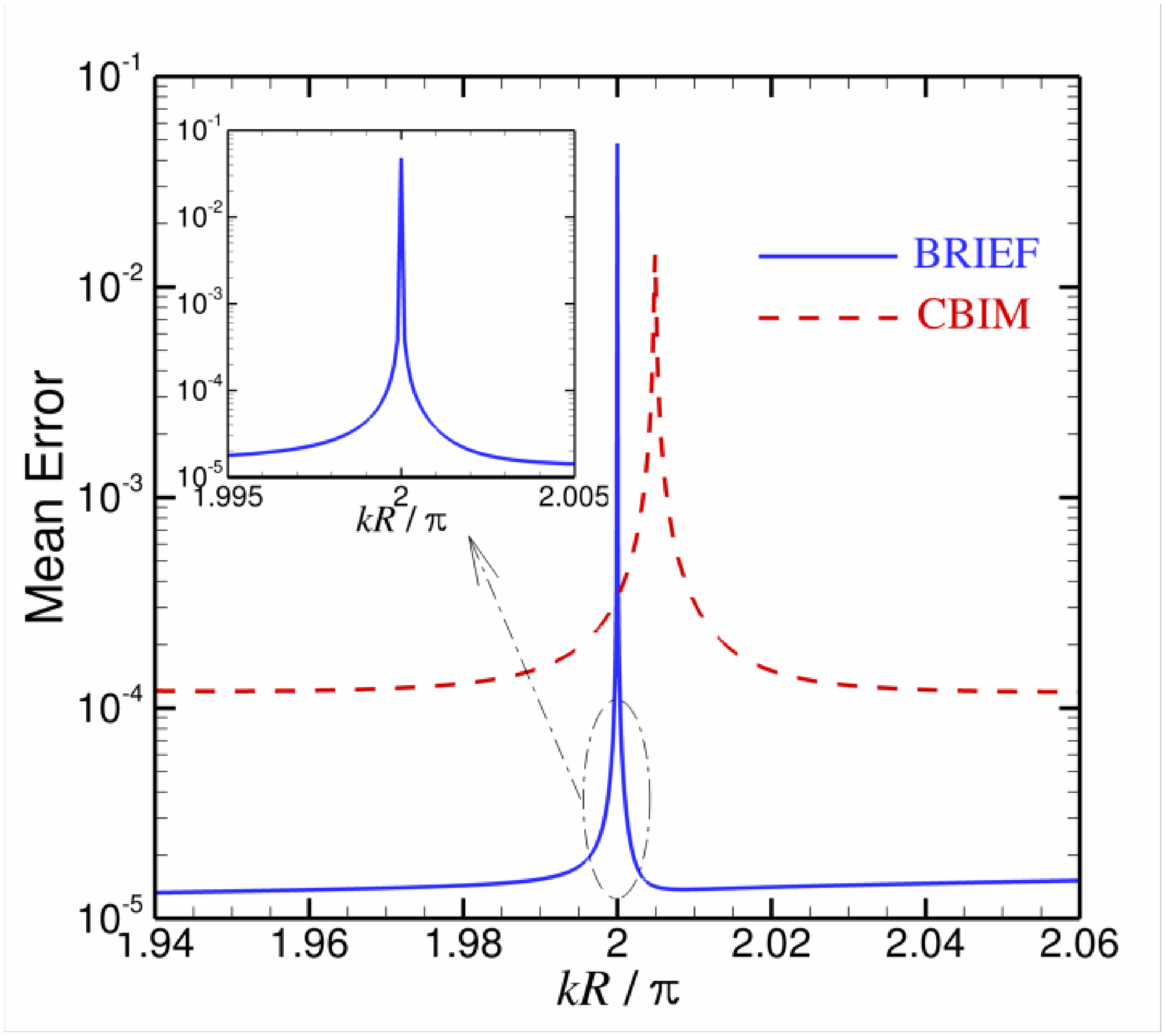} }
\caption{Comparison of the mean error defined in Eq.~\ref{eq:meanerror} as a function of the frequency near a resonant values (a) $k_f R = \pi$ and (b) $k_f R = 2\pi$ obtained using the conventional BEM (CBIM) approach and the desingularized BEM formulation (BRIEF). When using CBIM, the sphere surface is discretised with 2000 flat elements (DOF = 2000); while using BRIEF, the sphere surface is discretised with 980 quadratic elements connected by 1962 nodes (DOF = 1962). In the inset of (b), we see that the solution obtained using the BRIEF is unaffected by the fictitious solution when $kR$ is with 1 part in $10^4$ of $k_f R$.}
\label{fig:FigBRIEFvsCBIM}
\end{figure}

We now consider the example of solving the scattering problem by a solid sphere with radius $R$ for which the spectrum of resonant frequencies of the internal problem is known. A list of the values of the lower resonant frequencies and the equation that generates them are given in Table~\ref{tab:Table1} where we see that two of the lowest frequencies are at $k_f R=\pi$ and $k_f R=2\pi$. With the choice of Eq.~\ref{eq:Green1} for the Green's function, the spectrum fictitious frequencies coincides with the resonant frequencies of the corresponding internal problem. In Fig.~\ref{fig:FigBRIEFvsCBIM}, we quantify the behaviour of the BEM solution for $kR$ values in the neighborhood these 2 fictitious values in terms of the mean square error defined by
\begin{equation}\label{eq:meanerror}
     \text{Mean Error} = \frac{\sqrt{\sum_{i=1}^{\text{DOF}}\left(|\phi^{i}_{\text{num}}| - |\phi^{i}_{\text{ana}}|\right)^{2}}}{\text{DOF}},
 \end{equation}
where $\phi_{\text{num}}^i$ and $\phi_{\text{ana}}^i$ are, respectively, the numerical (BEM) and analytic solution at node $i$. The number of nodes used in the desingularized BEM, the Degree of Freedom (DOF), is around 2000.  We see that the mean squared error even in the small neighborhoods $0.94 \pi < kR < 1.06 \pi$ and $1.94 \pi < kR < 2.06 \pi$ around the 2 fictitious frequencies is extremely localized. In fact, the BEM solutions obtained by the desingularized BEM \cite{KlaseboerJFM2012, SunRoySoc2015} are unaffected by fictitious solutions until the frequency is within about 1 part in $10^4$ of a fictitious value. The results for the conventional boundary integral method (CBIM) are also shown. Note that the fictitious frequency predicted by the CBIM is significantly higher than the known theoretical value in these examples presumably because of the effect of approximating the sphere surface by a finite set of planar elements.

\begin{figure}[!t]
\centering\includegraphics[width=0.5\linewidth]{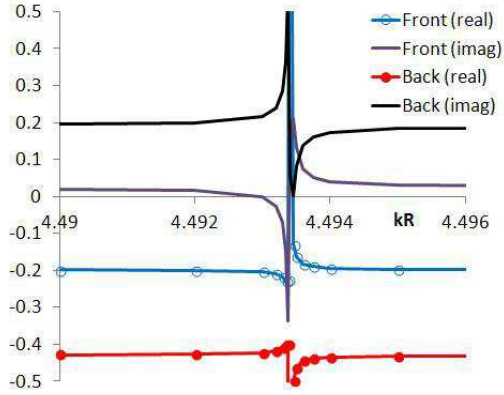}
\caption{Real and imaginary part of the scattered potential $\phi$ at the back and at the front of the sphere with the desingularized boundary element method showing the fictitious response around $k_fR=4.49341$, from $kR=4.490$ to $kR=4.496$. A quadratic mesh was used with 1442 nodes and 720 elements.}
\label{fig:FigSpikes4}
\end{figure}

Similar remarks apply for the behavior of the desingularized BEM solution in the neighborhood of the lowest $m=1$ fictitious value $k_f R=4.49341$ (see Table~\ref{tab:Table1}) shown Fig.~\ref{fig:FigSpikes4}. Here we show the values of the real and imaginary parts of the solution of nodes at the front and at the back of the sphere. The effect of the fictitious solution can only be discerned in the very narrow window $4.493 < kR < 4.494$ around $k_f R=4.49341$. But outside this window, there is no noticeable effect due to $kR$ being close to the fictitious value, $k_f R$. For example, if at the values $kR=3.140$ and $kR=4.490$ as given in Kinsler \cite{Kinsler}, page 518, the desingularized BEM (BRIEF) was used to to solve the Helmholtz equation, the solution would not register as giving fictitious results. As we shall see below, if a sweep of 10,000 frequencies from $kR=0$ to 10 is performed in steps of 0.001 one would miss many fictitious solutions (since a step size of 0.001 would not be precise enough to detect all of them). 

From the above results, we can conclude that the effects of resonance are not observed until one is extremely close to the resonant frequency in our desingularized BEM \cite{KlaseboerJFM2012, SunRoySoc2015}.

\section{The genesis of fictitious solutions} \label{sec:spuriousorigins}

In the example of acoustic scattering by a rigid scatterer that was discussed in the previous section, $\partial \phi / \partial n$ on the surface of the scatterer is specified (Neumann boundary conditions), and the variation of $\phi$ on the surface is the unknown to be found. At certain frequencies however, instead of the expected $\phi$, another function say, $\phi+f$ emerges. The frequencies at which this occurs, are often said to correspond to the internal resonance frequency of the same object. The Fredholm integral theory has been used to explain the occurrence of the fictitious frequency and related the fictitious frequency to the corresponding internal resonance frequency of the object \cite{Brascamp1969}.
However, by working directly with the integral equation that determines the fictitious solution, $f$, it is easy to demonstrate the origin of the fictitious solution. 

First we use the example of scattering on a rigid sphere of radius, $R$, to demonstrate how the fictitious solution and frequency is determined by the Green's function and the boundary shape. For simplicity, we consider the solution of the Helmholtz equation outside a sphere that has azimuthal symmetry for which the solution on the sphere surface can be expanded in terms of Legendre polynomials of order $m$, $P_m(\cos\theta)$ to account for variations in the polar angle, $\theta$. In this case, the fictitious frequencies for different $m$ values are known. We consider in detail the fictitious solution, $f$, and the fictitious frequency, $k_f$ for the cases with $m=0$ and $m=1$.
\subsection{Case: $f \sim P_0(\cos\theta), \text{a constant}$,  $m=0$}\label{sec:m0}
In this case, the fictitious solution, $f$ is a constant, being proportional to $P_0(\cos\theta)$, on the surface of the sphere of radius, $R$ and $c(\boldsymbol{x}_0)=2\pi$, then Eq.~\ref{eq:BEM2}, at the fictitious wave number, $k_f$, becomes:
\begin{equation} \label{eq:BEM3}
2 \pi+\int_S
\frac{\partial G(\boldsymbol x ,\boldsymbol x_0 | k_f)}{\partial n} \text{ d}S(\boldsymbol x) = 0.
\end{equation}
The integral of $\partial G(\boldsymbol x ,\boldsymbol x_0 | k)/\partial n$, can be evaluated (see \ref{sec:Appendix_A}) to give 
\begin{equation} \label{eq:integral_dG/dn}
\int_S \frac{\partial G(\boldsymbol x, \boldsymbol x_0 | k_f)}{\partial n} \text{ d}S(\boldsymbol x)= -2\pi\Big{\{}e^{i2k_f R}+\frac{1}{ik_f R}\big[1-e^{i2k_f R}\big] \Big{\}}
\end{equation}
so that Eq.~\ref{eq:BEM3} is equivalent to
\begin{equation} \label{eq:Resonance_m=0}
\sin(k_f R)[1-i k_f R]=0.
\end{equation}
Thus the spectrum of fictitious frequencies corresponding to a constant fictitious function, $f \sim P_0(\cos\theta)$, on the surface with $m=0$ is
\begin{equation} \label{eq:Spectrum_m=0}
\sin(k_f R) = 0 
\qquad \text{or} \qquad k_f R = \pi, 2\pi, 3\pi ....
\end{equation}
see also the first row of Table~\ref{tab:Table1}. In the external 3D domain, the fictitious solution $f(\boldsymbol x)$ that emerges numerically from the BEM solution corresponding to $k_f R=\pi$ is: $f(\boldsymbol x)=c_3 \; e^{ik_f\|\boldsymbol x\|}/\| \boldsymbol x\|$, where $c_3$ is an arbitrary constant and the origin of $\boldsymbol x$ taken at the origin of the sphere. 
\begin{table}[!t]
\centering
\caption{Values of the fictitious frequency that correspond to scattering by a rigid sphere with Neumann boundary condition. The three lowest values that are the solutions to the eigenvalue equation at each $m$ value given in the right most column are given to 6-7 significant figures.}
\label{tab:Table1}
\begin{tabular}{l l l l l}
\hline
$m$ &  & $k_f R$ &   & Equation: $x \equiv k_f R$ \\
 & $1^{st}$ & $2^{nd}$ & $3^{rd}$
\\
\hline
0 & 3.14159 & 6.28319 & 9.424778 & $\tan x=0$\\
1 & 4.49341 & 7.72525 & 10.90412 & $\tan x=x$\\
2 & 5.763459 & 9.095011 & 12.32294 & $\tan x=\frac{3x}{3-x^2}$\\
3 & 6.987932 & 10.41712 & 13.69802 & $\tan x =\frac{15x-x^3}{15-6x^2}$\\
4 & 8.182561 & 11.70491 & 15.03966 & $\tan x =\frac{105x-10x^3}{105-45x^2+x^4}$\\
5 & 9.355812 & 12.96653 & 16.35471  & $\tan x =\frac{945x-105x^3+x^5}{945-420x^2+15x^4}$\\
\hline
\end{tabular}
\end{table}
\subsection{Case: $f \sim P_1(\cos \theta)$,  $m=1$}\label{sec:m1}
A similar calculation to the one given in Section \ref{sec:m0}, for a fictitious function, $f \sim P_1(\cos\theta)$, for $m=1$ leads to (see \ref{sec:Appendix_B}) 
\begin{equation} \label{eq:tankr}
\tan(k_f R) = k_f R.    
\end{equation}
The first few solutions to Eq.~\ref{eq:tankr} are given in the $m=1$ row of Table~\ref{tab:Table1}. Again, these values are equal to those of the corresponding internal eigenvalue problem, yet they have been derived here purely from a boundary integral equation perspective.  Fictitious frequencies for higher order values of $m$ can also be obtained in a similar manner. Table~\ref{tab:Table1} contains all fictitious frequencies below $k_f R=10$ for a sphere. 

The above derivation that starts from the homogeneous integral equation, Eq.~\ref{eq:BEM2} demonstrates the role of the Green's function and the boundary shape in determining the spectrum of fictitious frequencies and solutions for acoustic scattering by a solid sphere. We can now show how to modify the fictitious frequency spectrum using different Green's functions. 
\section{The modified Green's function}
\label{sec:ModifiedGreen}
\begin{figure}[!t]
\centering\includegraphics[width=0.4\linewidth]{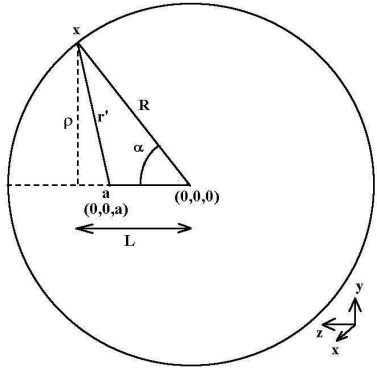}
\caption{Definition of the length $r'=\| \boldsymbol x -\boldsymbol a \|$, with $\boldsymbol a = (0,0,a)$ a fixed point inside the sphere with radius $R$. Also shown is the angle $\alpha$. The length $L$ satisfies $(L-a)^2+\rho^2=r'^2$ and since $\cos \alpha = L/R$, it follows that $a\cos \alpha = (R^2+a^2-r'^2)/(2R)$.}
\label{fig:FigDef2}
\end{figure}

Different forms of the Green's function can be used to construct the integral equation of the BEM as long as they satisfy the same differential equation in the solution domain and the Sommerfeld radiation condition at infinity as the free space Green's function. A simple modified Green's function, $G_{\text{mod}}$, can be taken as
\begin{equation}  \label{eq:G1}
\begin{aligned} 
    G_{\text{mod}}(\boldsymbol x, \boldsymbol x_0 | k) &\equiv G(\boldsymbol x, \boldsymbol x_0 | k) + \Delta G(\boldsymbol x, \boldsymbol x_0 | k) \\
      &= G(\boldsymbol x, \boldsymbol x_0 | k) + c_2 \; G(\boldsymbol x, \boldsymbol a | k)
\end{aligned}
\end{equation}
where the origin is taken to be the center of the sphere and the vector $\boldsymbol a$ corresponds to a point \emph{inside} the sphere $(|\boldsymbol a | < R)$ with $c_2$ an arbitrary constant. The integral equation that implements the BEM with $G_{\text{mod}}$ becomes: 
\begin{equation} \label{eq:BEM_Mod1}
\begin{aligned}
c\phi(\boldsymbol x_0)+  \int_S \phi(\boldsymbol x)
\Big[ & \frac{\partial G(\boldsymbol x , \boldsymbol x_0 | k)}{\partial n} + c_2\frac{\partial  G(\boldsymbol x ,\boldsymbol a | k)}
{\partial n} \Big]\text{ d}S(\boldsymbol x) \\ 
& \qquad  =  \int_S \frac{\partial \phi(\boldsymbol x)}{\partial n} \big[G (\boldsymbol x, \boldsymbol x_0 | k)+c_2 G (\boldsymbol x, \boldsymbol a | k) \big]\text{ d}S(\boldsymbol x).
\end{aligned}
\end{equation}
The additional term $G(\boldsymbol x, \boldsymbol a|k)$ although singular at the location $\boldsymbol a$, does not create any singular behavior on the surface $S$, since $\|\boldsymbol x - \boldsymbol a\|$ never becomes zero (see also Fig.~\ref{fig:FigDef2}). The modified Green's function, $G_{\text{mod}}(\boldsymbol x, \boldsymbol x_0 | k)$, also satisfies the Sommerfeld radiation condition at infinity.
\subsection{Case: $f \sim P_0(\cos\theta), \text{a constant}$,  $m=0$ with modified $G_{\text{mod}}$} \label{sec:ModifiedGreen_m0}
\begin{figure}[!t]
\centering
\subfloat[]{\includegraphics[width=0.45\linewidth]{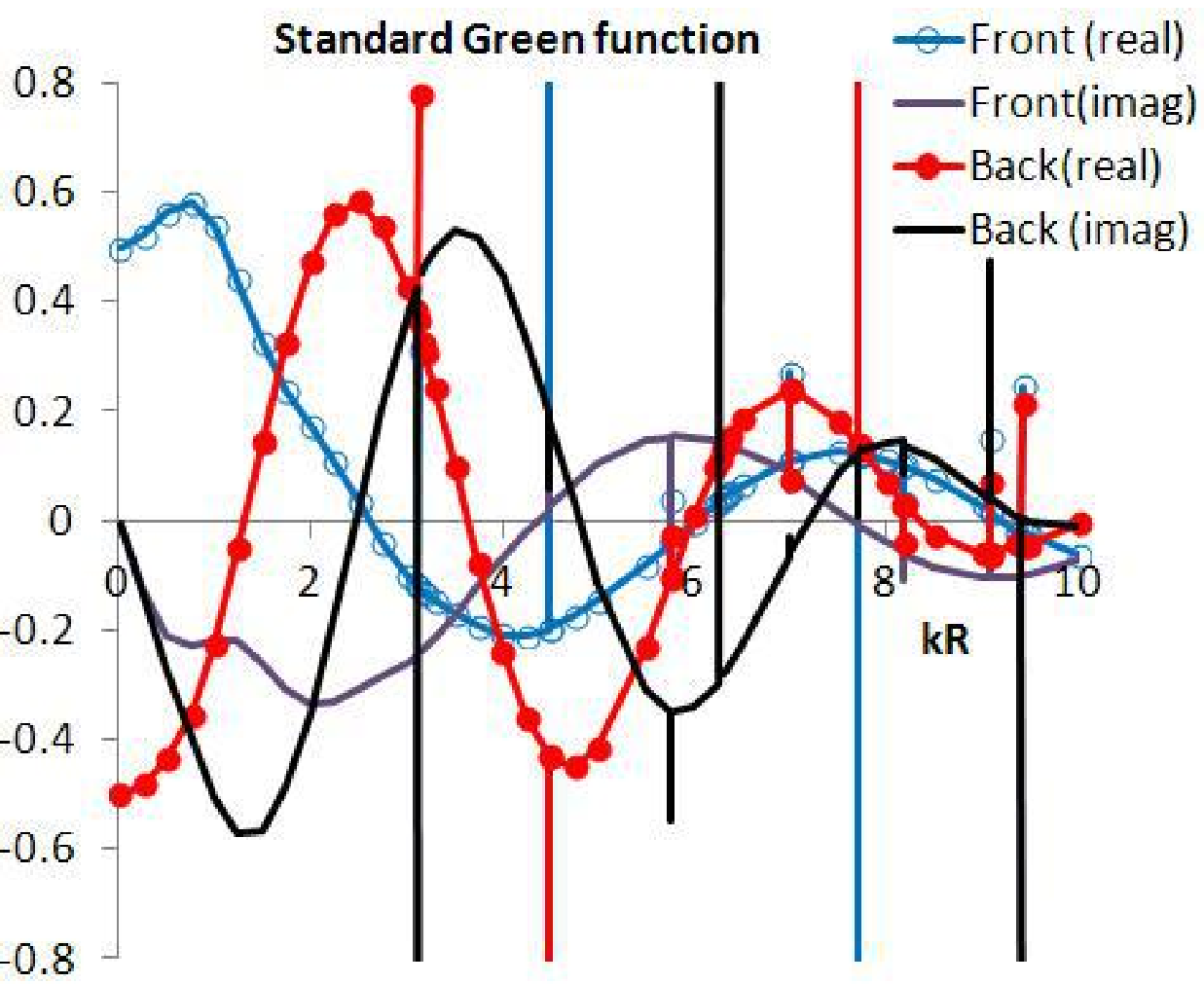} }
\subfloat[]{\includegraphics[width=0.45\linewidth]{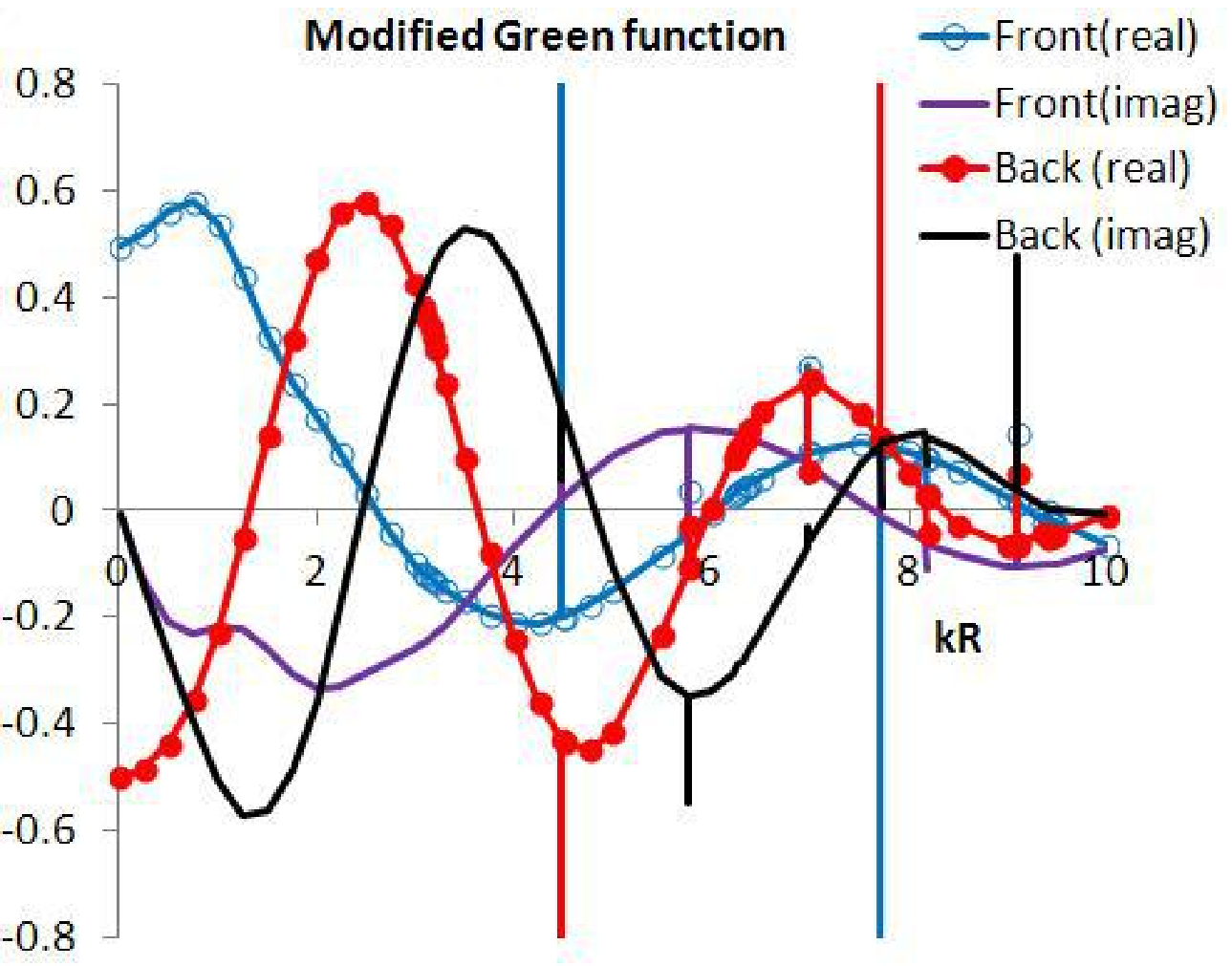} }
\caption{(a) Results obtained with the desingularized boundary element method \cite{KlaseboerJFM2012, SunRoySoc2015} with a classical free space Green's function, Eq.~\ref{eq:Green1}, with 720 six node quadratic elements and 1442 nodes. The real and imaginary part of the scattered $\phi$ in front of and behind a sphere with radius $R$ due to an incident plane wave with wavenumber $k$ as a function of $kR$. The effect of fictitious solutions can clearly be observed as sharp peaks and correspond to fictitious frequencies listed in Table \ref{tab:Table1}. More data points have been used near the fictitious frequencies. (b)
Results using the modified Green's function, Eq.~\ref{eq:G1}. The fictitious responses corresponding to $ka=\pi$, $ka=2\pi$ and $ka=3\pi$ are now eliminated. Besides the implementation of the modified Green's function, the parameters used are the same as those in a).}
\label{fig:FigSpikes1}
\end{figure}

Let us now investigate how the modified Green's function defined in Eq.~\ref{eq:G1} and \ref{eq:BEM_Mod1} can affect the spectrum of fictitious frequencies that is now determined by
\begin{equation} \label{eq:Homog_BEM_G1}
2 \pi+\int_S
\Big[ \frac{\partial G(\boldsymbol x ,\boldsymbol x_0 | k_f)}{\partial n} + \frac{\partial G(\boldsymbol x ,\boldsymbol a | k_f)}{\partial n} \Big] \text{ d}S(\boldsymbol x) = 0.
\end{equation}
Evaluating the integrals (see \ref{sec:Appendix_C}) then gives the equation that determines the spectrum of fictitious frequencies
\begin{equation} \label{eq:BEM_Mod9}
\sin(k_f R)+ c_2 (R/a) \sin(k_f a)=0.
\end{equation}
Thus the original fictitious frequency spectrum given by $\sin(k_f R)=0$ in Eq.~\ref{eq:Spectrum_m=0} due to the use of the unmodified Green's function in Fig.~\ref{fig:FigSpikes1}a has been replaced by a  different spectrum given by Eq.~\ref{eq:BEM_Mod9} in Fig.~\ref{fig:FigSpikes1}b. Furthermore, the precise value of $a$ is not critical. In fact, since $\sin(k_f a)/a \rightarrow k_f$ as $a \rightarrow 0$, we can put $\boldsymbol a = \boldsymbol 0$, that is, at the center of the sphere. In the results shown in Fig.~\ref{fig:FigSpikes1}, we have taken $a=0$ and $c_2=-1$.



%

For $m=0$, Eq.~\ref{eq:BEM_Mod9} will in general assure that the new fictitious frequency spectrum obtained with the modified Green's function, $G_{\text{mod}}$ will be different from that obtained with the original Green's function, $G$. However, there are still ways for which this may not be true. 
\begin{figure}[!t]
\centering\includegraphics[width=0.5\linewidth]{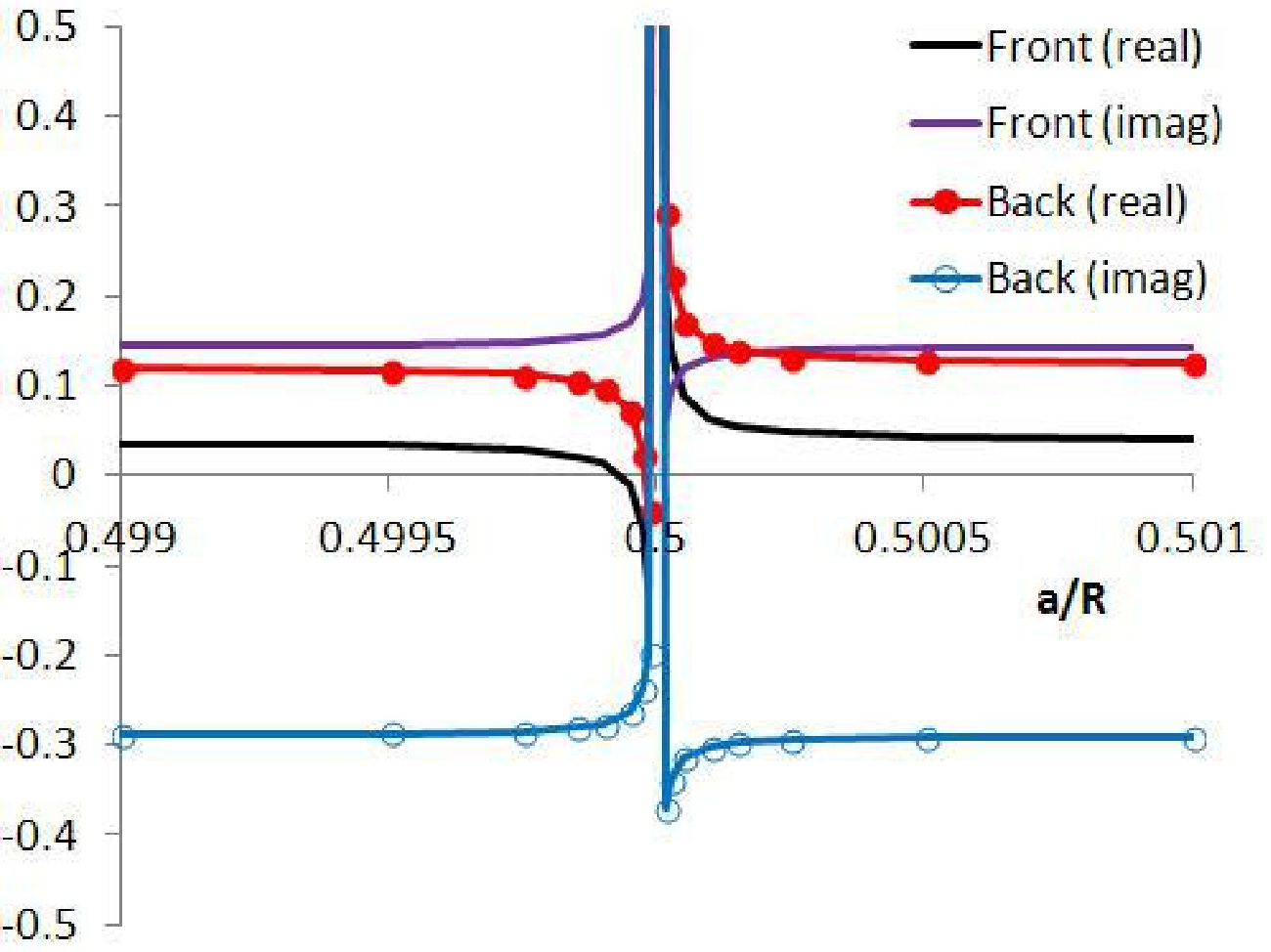}
\caption{An example where the original and modified spectrum have common values. Here $kR=2\pi$ is fixed and $a/R$ is varied slightly near the value $a/R=0.5$, (thus $ka=\pi$) resulting in $\sin(k_fR)=0$ and $\sin(k_fa)=0$ simultaneously in Eq.~\ref{eq:BEM_Mod9} and the modified Green's function framework fails. Plotted are the real and imaginary part of the scattered potential $\phi$ at the nodes in front and at the back of the sphere. In the neighbourhood of $a/R=0.5$, the solution is still accurate up to 2\% at $a/R=0.499$ and $a/R=0.501$. The value at $a/R=0.5$ is highly erroneous at $\phi_{Front}=1.51+i5.72$ and $\phi_{Back}=1.60+i5.29$ (for $a=0$, $\phi_{Front}=0.03858+i0.1443$ and $\phi_{Back}=0.1230-i0.2893$). A quadratic mesh was used with 1442 nodes and 720 elements. }
\label{fig:Figa05}
\end{figure}
\begin{itemize}
\item Firstly, it is still possible that both $\sin(k_fR)$ and $\sin(k_fa)$ vanish, that is, the original spectrum and the modified spectrum contain common values. An example of such a case can be observed when $k_fR=2\pi$ and $a=0.5R$ (thus $k_fa=\pi$ and $\sin(k_fa)=0$). This was tested numerically and indeed for these parameters there is still a spurious solution corresponding to the common fictitious frequency values in the 2 spectra as illustrated in Fig.~\ref{fig:Figa05}.

%
\item A second way in which fictitious behaviour can still be observed, is when for particular parameters of $k_f$, $R$, $a$ and $c_2$, Eq.~\ref{eq:BEM_Mod9} is still zero. An instance of such fictitious behavior can be observed for the parameters $k_fR=0.5$, $a=0.3R$ and $c_2=-0.9624563$. The fictitious solution for these parameters is about 100 times the theoretical value in a numerical test. It is interesting to note that a fictitious frequency now appears at $k_fR=0.5$, a frequency value that was previously free of fictitious behavior.  This is an example of a frequency shift of the lowest fictitious behavior from $k_fR=\pi$ to a lower frequency of $k_fR=0.5$. However, if $c_2=-0.9620000$ is chosen, thus only slightly different from $c_2=-0.9624563$, no fictitious behavior is observed at all (see Fig.~\ref{fig:Fig_c2}). 
\item Finally, the location of the point $\boldsymbol a$ should not be chosen too close to the boundary $S$. In order to investigate this, in Fig.~\ref{fig:Fig_a_to_R}, the potentials in front and at the back of the sphere are shown, while the location of $\boldsymbol a$ of the modified Green's function is varied from $a=0.0$ to $1.0$. From the figure it can be deduced that $\boldsymbol a$ should not be placed closer to the boundary $S$ than roughly the meshsize.
\end{itemize} 
\begin{figure}[!t]
\centering\includegraphics[width=0.5\linewidth]{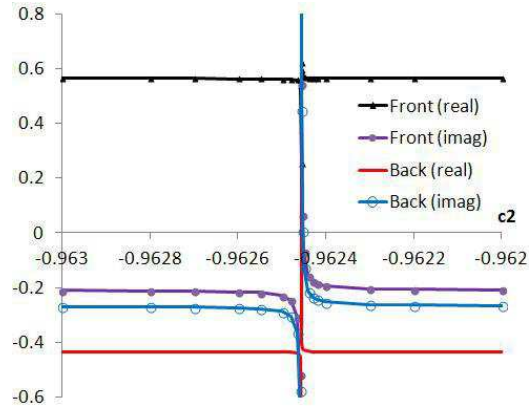}
\caption{Fictitious behavior when for particular parameters of $k_f$, $R$, $a$ and $c_2$, Eq.~\ref{eq:BEM_Mod9} is still zero. Here we have the case $k_fR=0.5$ and $a=0.3R$ and the parameter $c_2$ is varied from $-0.963$ to $-0.962$. Only when $c_2$ is very close to the "critical" value of $c_2=-0.9624563$ does the solution starts to degenerate. The value at $c_2=-0.9624563$ has large errors at $\phi_{Front}=0.2531-i8.352$ and $\phi_{Back}=-0.7462-i8.410$. These results were obtained with a quadratic mesh with 1442 nodes and 720 elements.}
\label{fig:Fig_c2}
\end{figure}

To conclude, for $m=0$, the modified Green's function approach can indeed remove the fictitious behavior of the solution. In the next section, the $m=1$ case will be investigated. 

\begin{figure}[!t]
\centering\includegraphics[width=0.5\linewidth]{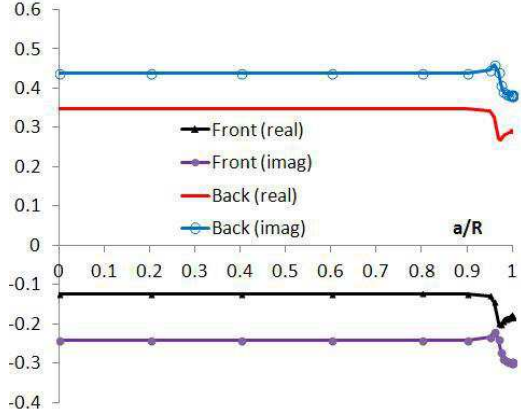}
\caption{Variation of the potentials $\phi$ in front and at the back of the sphere as a function of $a/R$, the parameter $\boldsymbol a = (a,0,0)$ in the modified Green's function with $kR=\pi$ and $c_2=1.0$. The results were obtained with a quadratic mesh with 1442 nodes and 720 elements, which results in an average distance between nodes of about $0.05R$. This is roughly the distance where the solution starts to deviate from the analytical value at $a/R=0.95$. The solution does not diverge, even at exactly $a=R$, although the value is incorrect.}
\label{fig:Fig_a_to_R}
\end{figure}

\subsection{The $m=1$ case} \label{sec:ModifiedGreen_m1}

In Section \ref{sec:ModifiedGreen_m0}, it was shown that for $f=$constant (or $m=0$), the modified Green's function can indeed remove the fictitious solutions. A similar proof can now be attempted for $m=1$. In analogy to Eq. \ref{eq:App1}, it must now be shown that
\begin{equation}
\label{eq:AppB1}
2\pi R+\int_S z\Big[
\frac{\partial G(\boldsymbol x ,\boldsymbol x_0|k_f)}
{\partial n} +c_3\frac{\partial G(\boldsymbol x ,\boldsymbol a|k_f)}
{\partial n}\Big]\text{ d}S(\boldsymbol{x}) = 0.    
\end{equation}
Thus the integral 
\begin{equation}
\label{eq:AppB2}
\int_S z
\frac{\partial G(\boldsymbol x ,\boldsymbol a|k_f)}
{\partial n}\text{ d}S(\boldsymbol{x})    
\end{equation}
must be determined. The framework of Eqs. \ref{eq:BEM_Mod5}, \ref{eq:BEM_Mod6} can be adapted immediately, provided that we add $z$ in the equations. With $z=R \cos \alpha = [R^2+a^2-r'^2]/(2a)$:
\begin{equation}
\begin{aligned}
\label{eq:BEM_AppB3}
\int_S
z\frac{\partial G(\boldsymbol x ,\boldsymbol a|k_f)}
{\partial n} \text{ d}S(\boldsymbol{x}) = \qquad \qquad \qquad \qquad \\   \frac{2\pi R}{a}\int_{R-a}^{R+a} 
\frac{R^2+a^2-r'^2}{2a} 
\Big[-R + \frac{R^2+a^2-r'^2}{2R} \Big]\frac{e^{ikr'}}{r'^2}[ikr'-1] \text{ d}r'
\end{aligned}
\end{equation}

This integral can be shown not to be equal to zero. However, a similar calculation for $x$ or $y$ instead of $z$, shows that due to symmetry (provided that $\boldsymbol x_0$ is still situated on the z-axis):

%
\begin{equation}
\label{eq:AppB4}
\int_S x
\frac{\partial G(\boldsymbol x ,\boldsymbol a|k_f)}
{\partial n}\text{ d}S(\boldsymbol{x}) =
\int_S y
\frac{\partial G(\boldsymbol x ,\boldsymbol a|k_f)}
{\partial n}\text{ d}S(\boldsymbol{x})= 0    
\end{equation}
From this we can conclude that, unfortunately, the fictitious solutions corresponding to $m=1$ cannot be removed when applying our modified Green's function in its present form. This is also clear from Fig.~\ref{fig:FigSpikes1}b, the fictitious behavior corresponding to $m=1$ is still present. A more elaborate Green's function might still be capable of removing these frequencies as well, but this is beyond the scope of the current article, in which we intend merely to show the proof of concept.

\section{Discussion}\label{sec:Discussion}

\begin{figure}[!t]
\centering
\subfloat[]{ \includegraphics[width=0.4\linewidth]{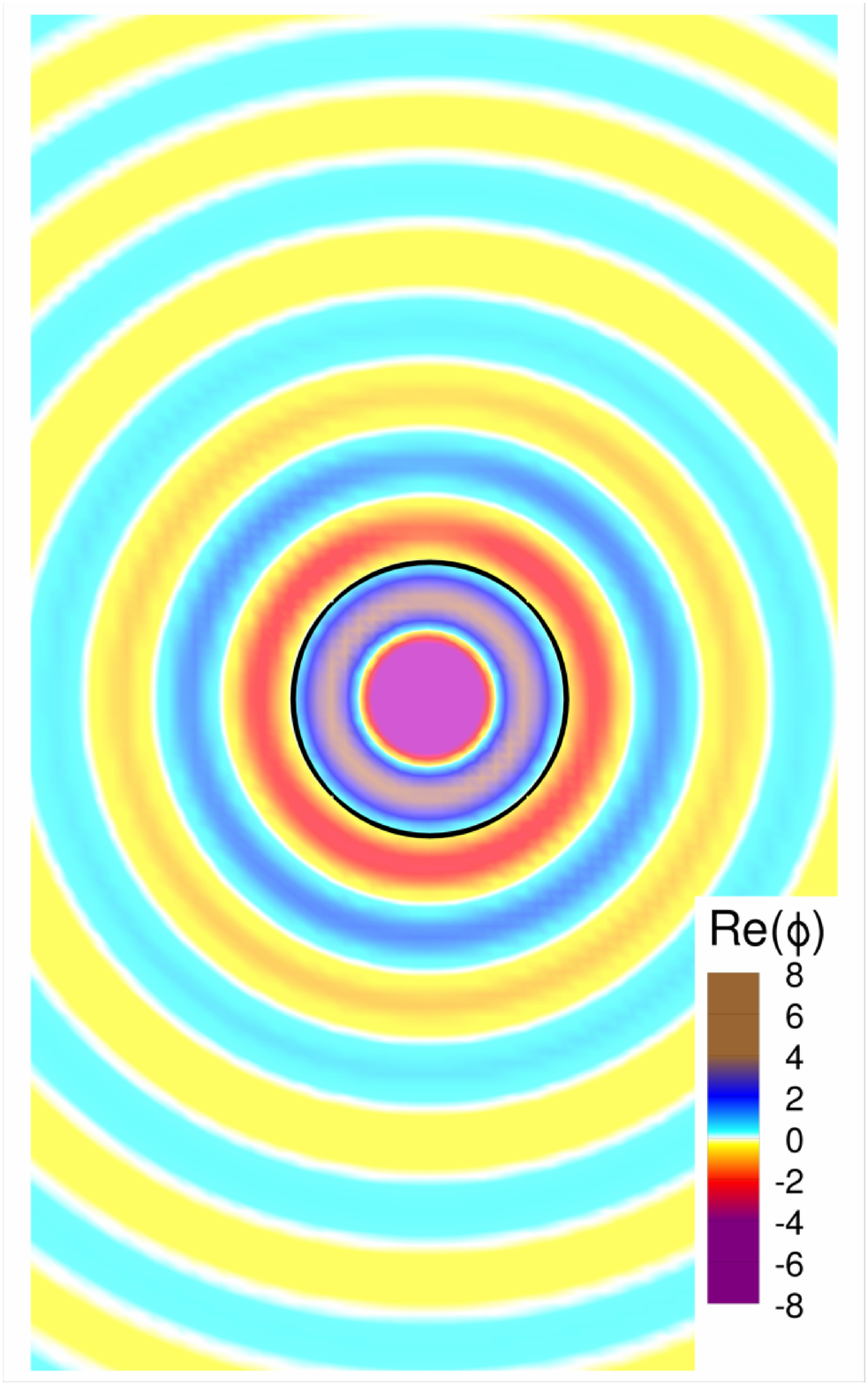} }
\subfloat[]{ \includegraphics[width=0.4\linewidth]{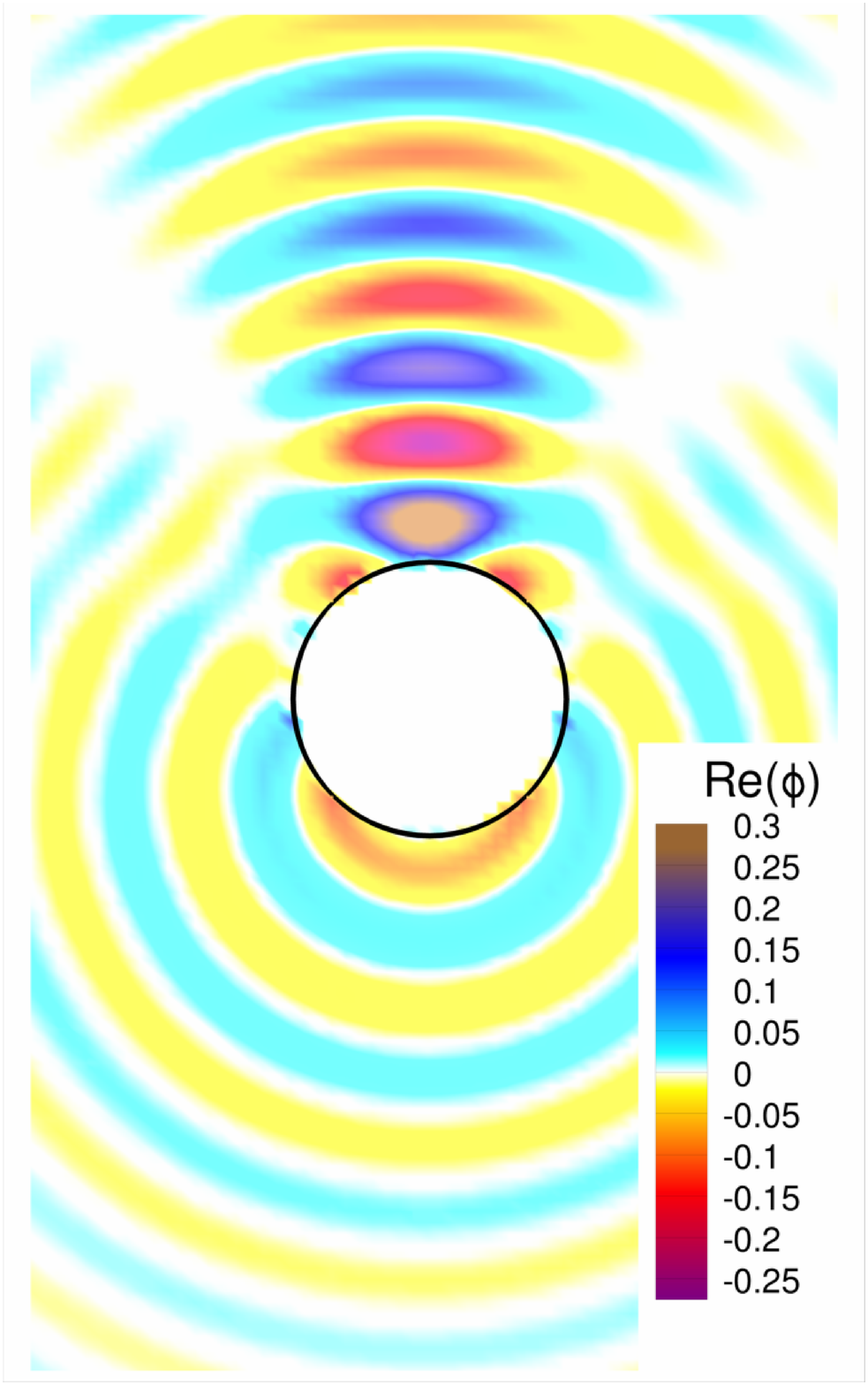} }
\caption{Field plot of the real part of the potential $\phi$ obtained with Eq.~\ref{eq:BEM_Field} for $kR=2\pi$: (a) with the standard (desingularized) BEM method where fictitious results are present and the fictitious solution inside the sphere (indicated by a black circle) can clearly be observed; (b) with the modified Green's function, no resonance solution is visible, the solution inside the sphere is very close to zero.} 
\label{fig:phi2pi}
\end{figure}

In both the modified Green's function and in the CHIEF method, a point in the interior of the domain is chosen on which an integral equation for $G(\boldsymbol x, \boldsymbol a)$ is developed. The difference between the modified Green's function and CHIEF, however, lies in the fact that CHIEF uses the following equation as an extra condition to the system of equations:
\begin{equation} \label{eq:BEM_CHIEF}
\begin{aligned}
c\phi(\boldsymbol a)+ \int_S \phi(\boldsymbol x)
\frac{\partial  G(\boldsymbol x ,\boldsymbol a|k)}
{\partial n} \text{ d}S = \int_S \frac{\partial \phi(\boldsymbol x)}{\partial n} G (\boldsymbol x, \boldsymbol a|k) \text{ d}S,
\end{aligned}
\end{equation}
Here, the constant $c=0$, since the point $\boldsymbol a$ is situated outside the domain (i.e. inside the object) in the CHIEF method. In the modified Green's function approach this equation is essentially added to the 'normal' Green's function.

A way to check if the solution using our desingularized boundary element code for a general shaped object contains a fictitious component due to $k$ being close to a fictitious value is to repeat the  calculation at a very slightly different $k$ value. If the solution differs significantly, the solution is likely to contain a fictitious component. 

%

We further illustrate the concepts with some field values of $\phi$ obtained by post-processing from the following equation
\begin{equation} \label{eq:BEM_Field}
\begin{aligned}
4 \pi\phi(\boldsymbol x_0)= -\int_S \phi(\boldsymbol x)
\frac{\partial  G(\boldsymbol x ,\boldsymbol x_0|k)}
{\partial n} \text{ d}S + \int_S \frac{\partial \phi(\boldsymbol x)}{\partial n} G (\boldsymbol x, \boldsymbol x_0|k) \text{ d}S,
\end{aligned}
\end{equation}
where $x_0$ is not situated on the boundary $S$, but either in the solution domain or inside the sphere (outside the solution domain). If no resonance is present, the solution inside the sphere (and hence outside the solution domain) should be $\phi=0$. In for following examples we use 1442 nodes and 720 quadratic elements in the BEM solution. The first case is the solution for $kR=2\pi$ where in Fig.~\ref{fig:phi2pi} we plotted the results obtained from both the standard BEM (with fictitious results, Fig.~\ref{fig:phi2pi}a and that obtained using a modified Green's function Fig.~\ref{fig:phi2pi}b, where the solution inside the sphere is zero.

\begin{figure}[!t]
\centering
\subfloat[]{ \includegraphics[width=0.4\linewidth]{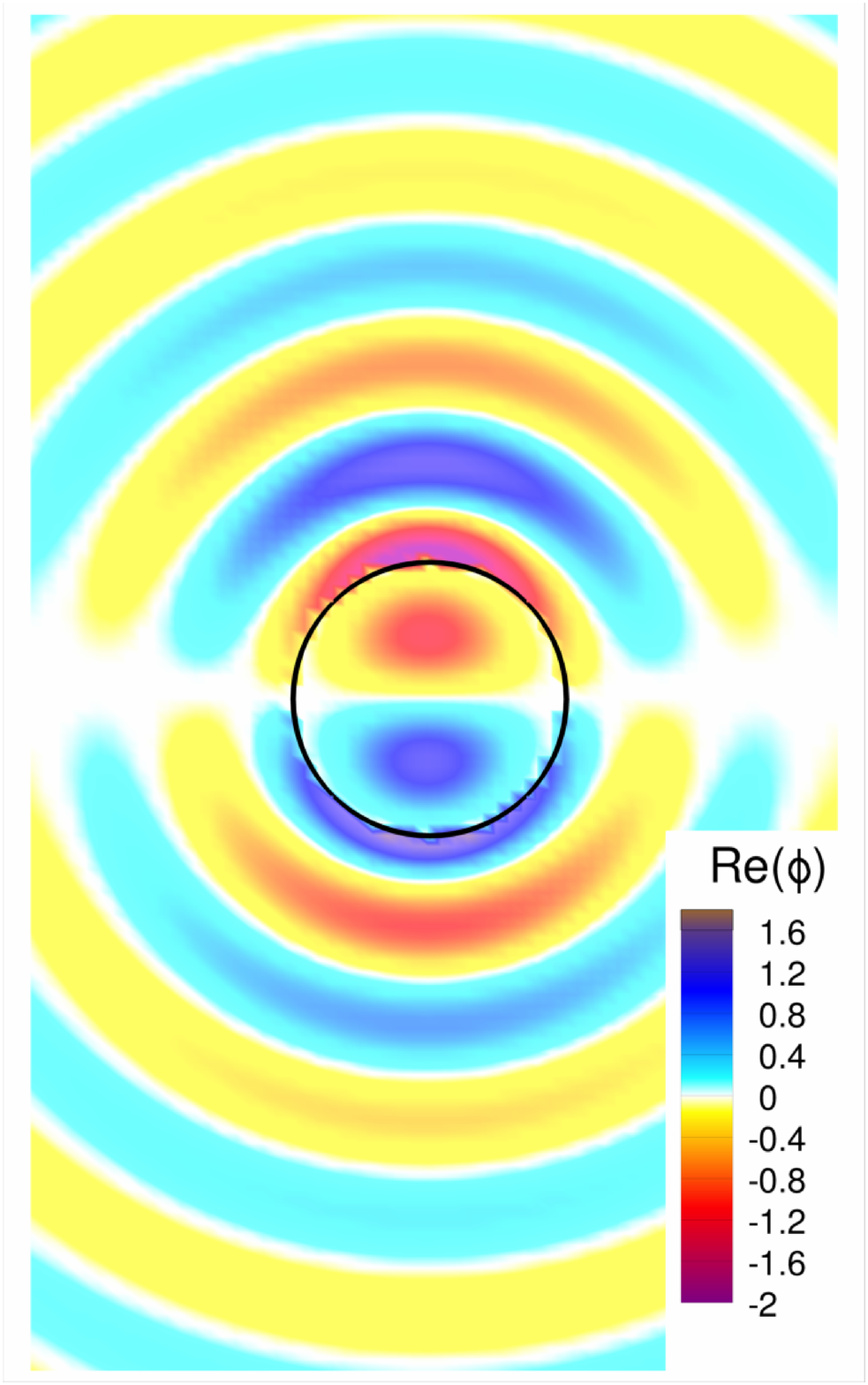} }
\subfloat[]{ \includegraphics[width=0.4\linewidth]{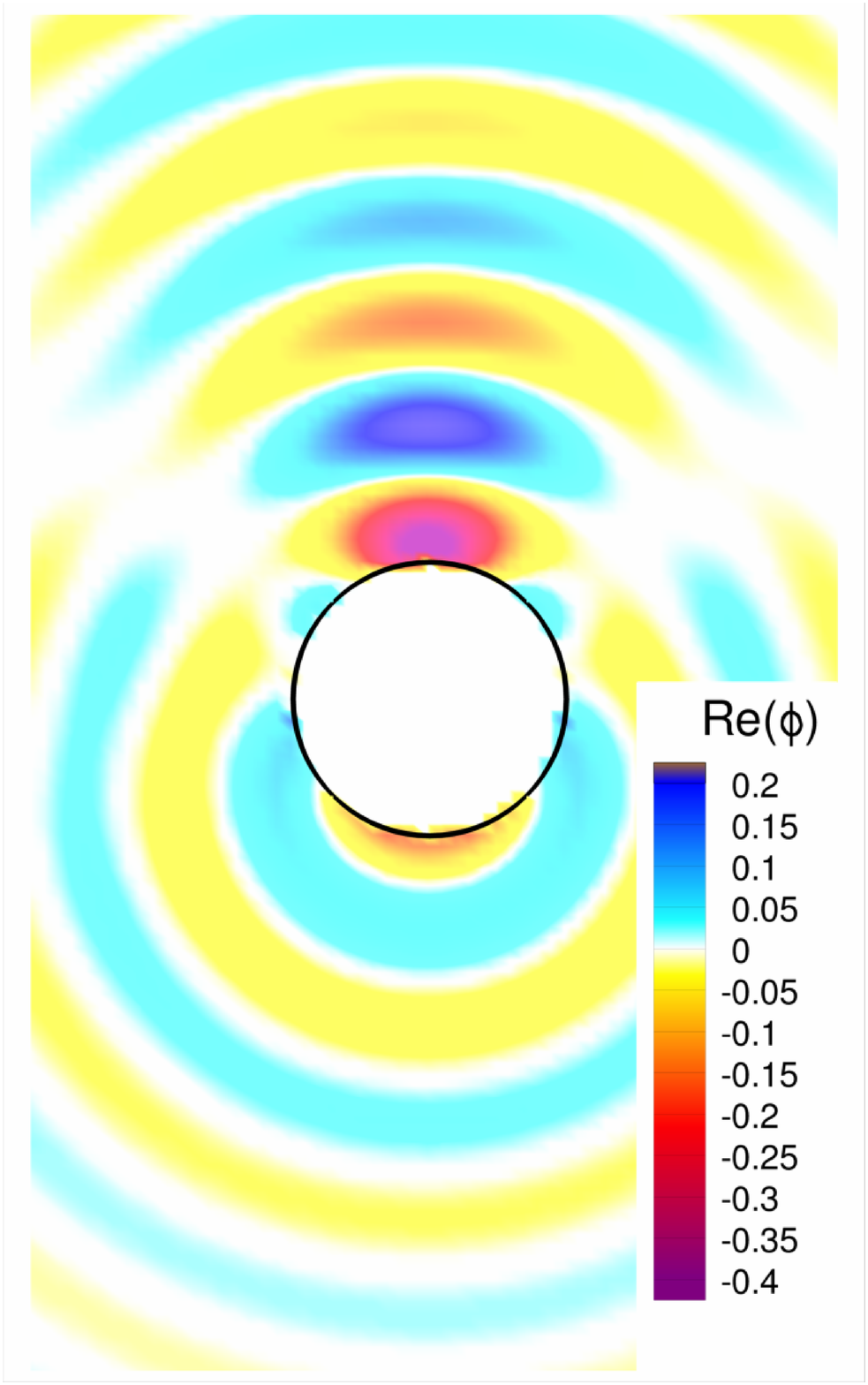} }
\caption{Field plot of the potential (real part) for (a) the resonance frequency $kR=4.49341$ and (b) near this frequency at $kR=4.49000$; both with the standard (desingularized) BEM. In (a) the fictitious solution can clearly be seen inside the sphere. No resonance solution is visible in (b), the solution inside the sphere is zero. The plots emphasize the superior accuracy of the desingularized BEM: if the frequency is only slightly besides a resonance value, the desingularized BEM still gives the correct result.}
\label{fig:phi4_49}
\end{figure}

A second example shows the solution for the resonance frequency $kR=4.49341$ in Fig.~\ref{fig:phi4_49}a. At a frequency nearby at $kR=4.49000$ no resonance behavior is observed in Fig.~\ref{fig:phi4_49}b. This once more demonstrates the extreme accuracy of our desingularized BEM framework.

\begin{figure}[!t]
\centering
\subfloat[]{ \includegraphics[width=0.4\linewidth]{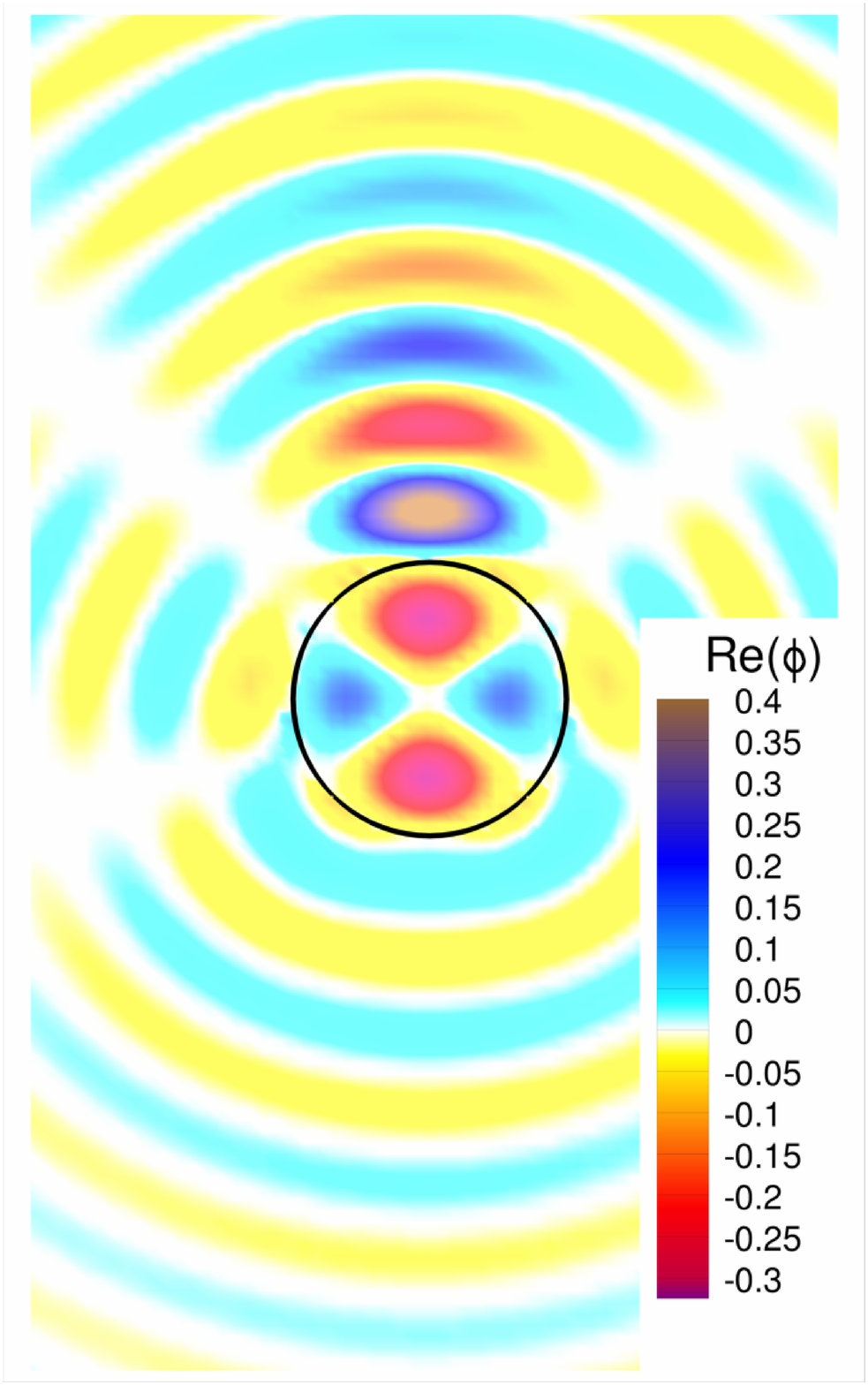} }
\subfloat[]{ \includegraphics[width=0.4\linewidth]{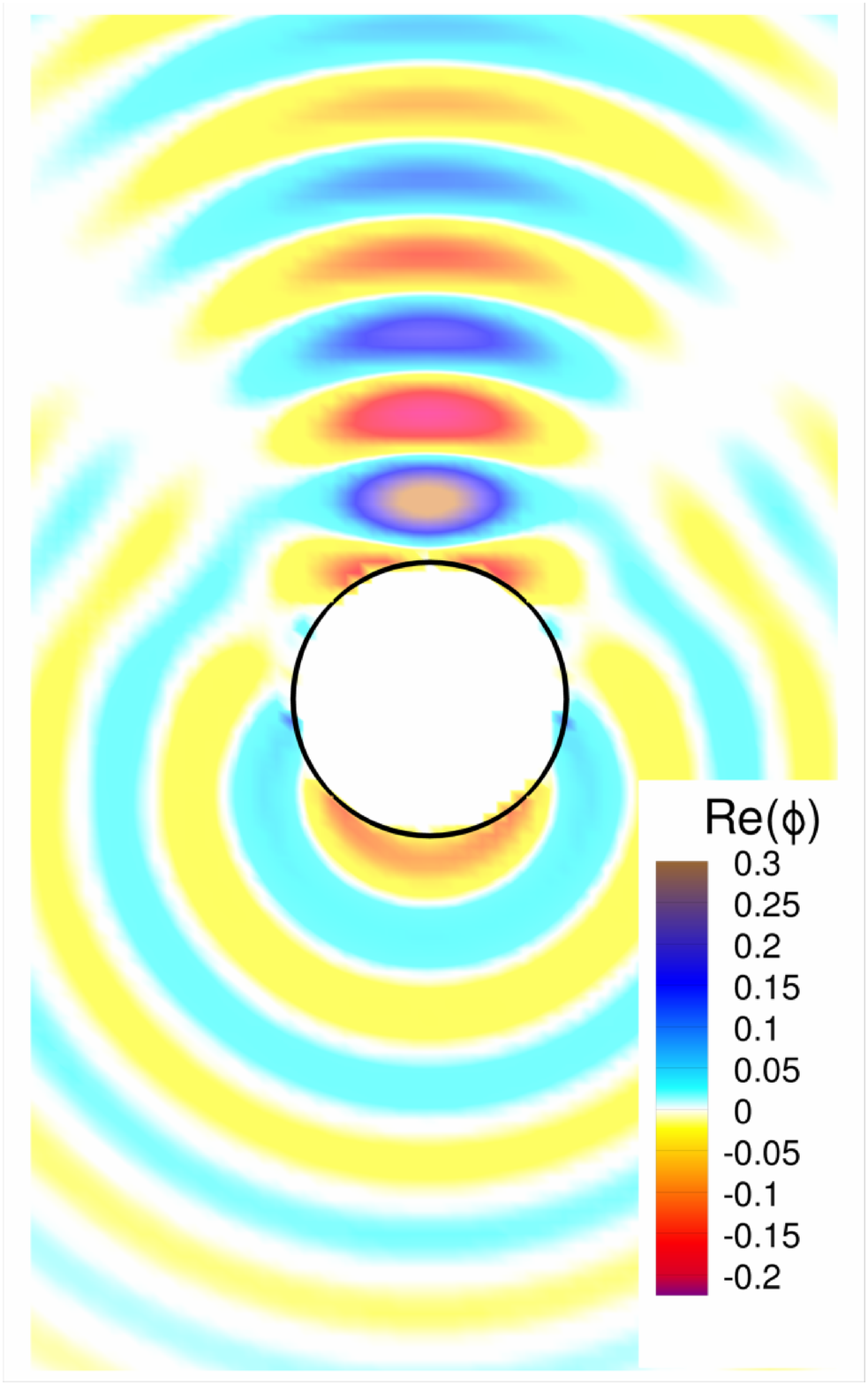} }
\caption{Potential (real) plot obtained by post processing for (a) $kR=5.76345$ and (b) $kR=5.76000$; both with the standard (desingularized) BEM. In (a) the fictitious solution can clearly be observed inside the sphere. No resonance solution can be observed in (b). The plots again emphasize the superior accuracy of the desingularized BEM and also show a graphical means to test if the solution exhibits fictitious behavior or not. }
\label{fig:phi5_76}
\end{figure}

A third example shows the resonance behavior at $kR=5.76345$ and a nearby value of $kR=5.76000$ in Fig.~\ref{fig:phi5_76}. Again no resonant behavior is observed at the nearby value. 

\begin{figure}[!t]
\centering
\subfloat[]{ \includegraphics[width=0.4\linewidth]{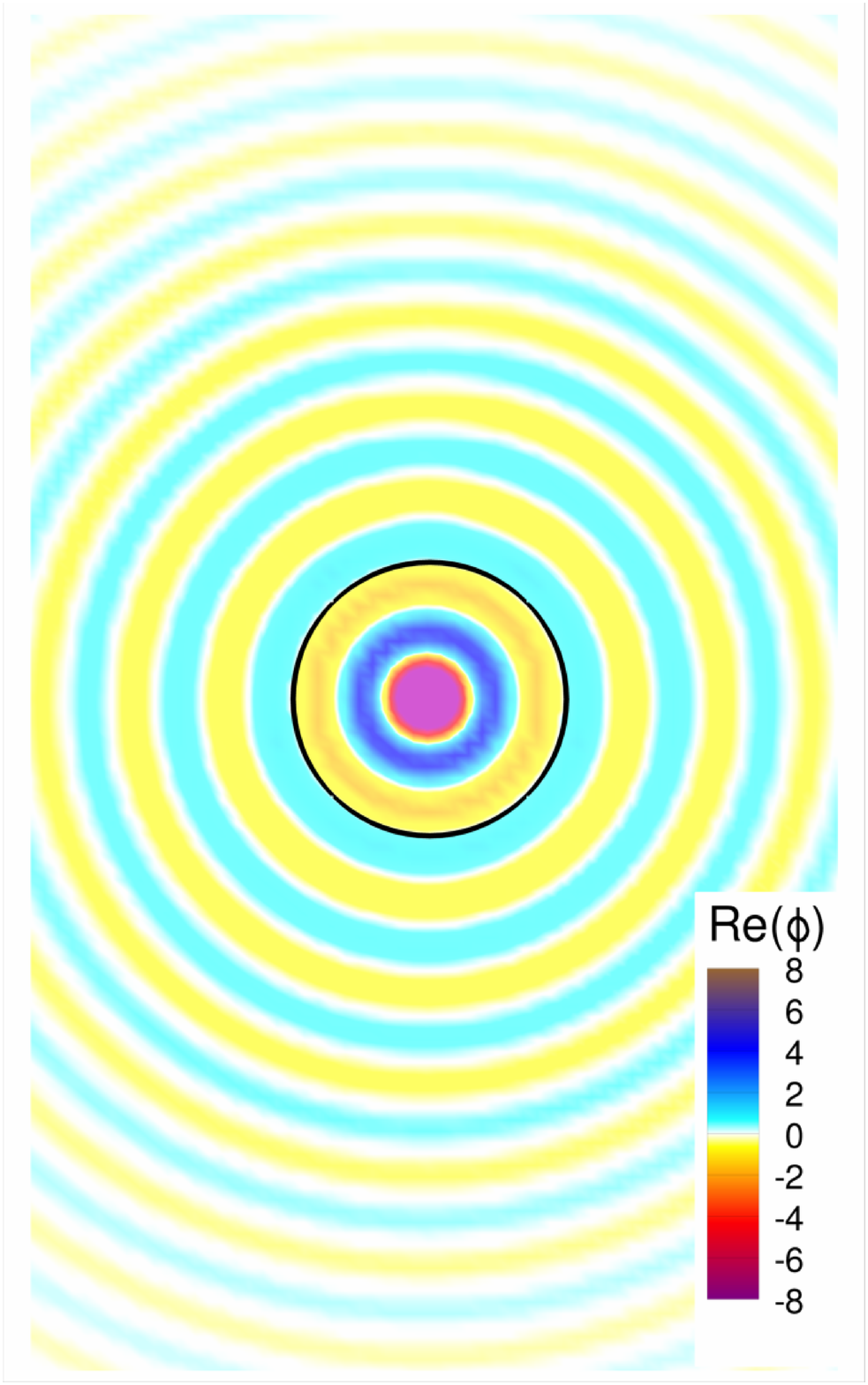} }
\subfloat[]{ \includegraphics[width=0.4\linewidth]{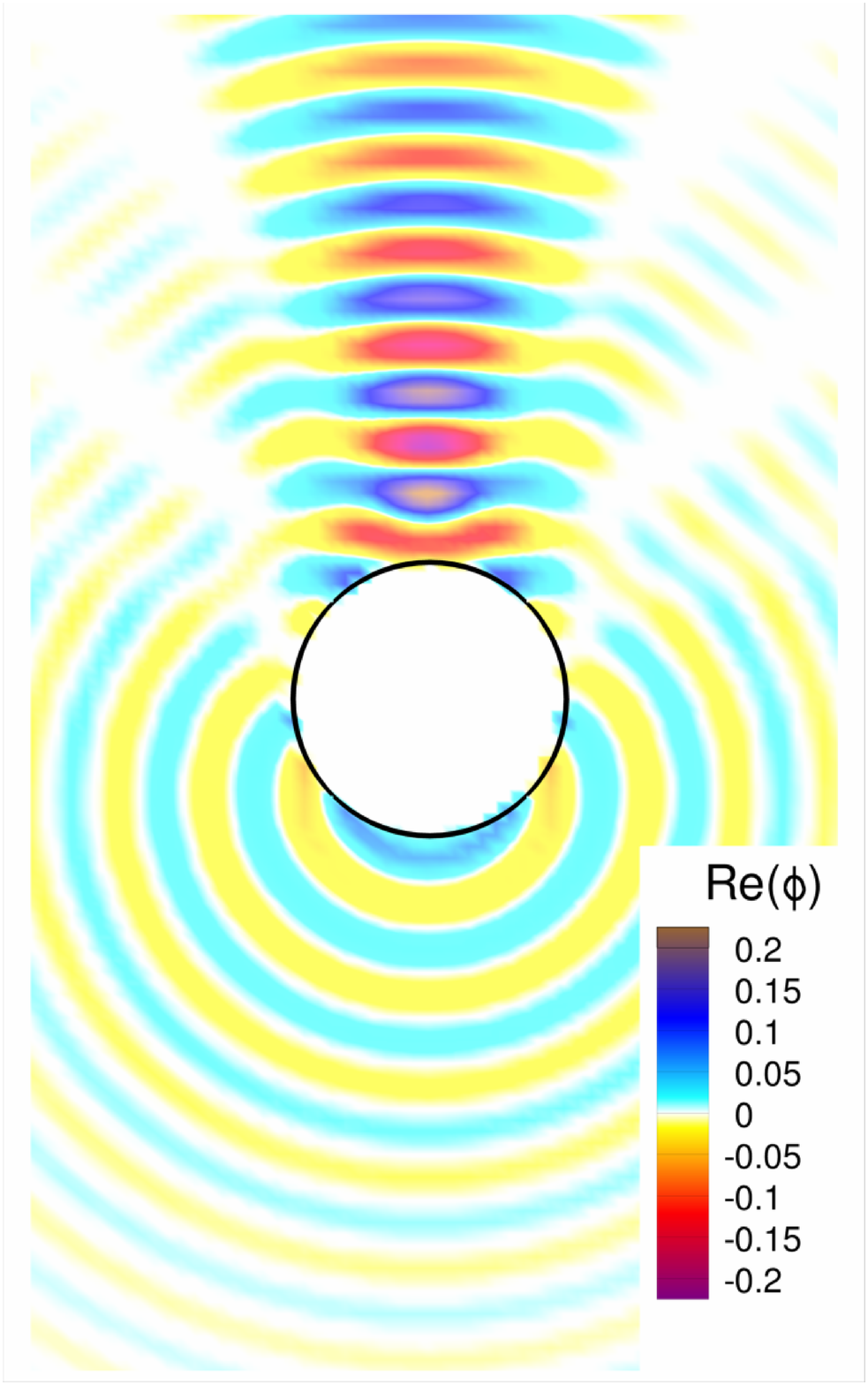} }
\centering
\caption{Plot of the real part of the potential obtained by post processing for (a) $kR=3\pi$ (with the standard, desingularized method) and (b) $kR=3\pi$ with the modified Green's function (desingularized as well). The fictitious spherical symmetrical solution inside the sphere in (a), which totally overshadows the real solution has successfully been eliminated in (b).}
\label{fig:phi3pi}
\end{figure}

A final example shows the solution at $kR=3\pi$ in Fig.~\ref{fig:phi3pi}, obtained from both with the standard method and with the modified Green's function. 

At present, the modified Green's function can only remove solutions associated with fictitious frequencies in the ``breathing modes'' ($m=0$), but these are most likely the first modes to appear with increasing $k$. It would be interesting to find other modified Green's functions to remove  solutions associated with fictitious frequencies in all modes, but we have not as yet been able to develop such an approach. 

\section{Conclusions}\label{sec:Conclusions}

The fictitious frequencies occurring in a BEM implementation of the Helmholtz equation were revisited. From a BEM viewpoint it was highlighted how these fictitious solutions appear and how they can be detected. It was shown that the use of a modified Green's function can indeed remove certain fictitious frequencies. To the best knowledge of the authors, this is the first time actual numerical results have been obtained with a modified Green's function. The results presented are a demonstration of the proof of concept. More elaborate modified Green's functions might be able to remove more fictitious frequencies. If indeed so, then this easy to implement method could be a viable alternative to existing methods.

Fictitious frequencies cannot fully be avoided with the current alternative Green's function approach, but it is sometimes possible to shift this frequency to another region of the spectrum. Thus the fictitious frequencies do not necessarily coincide anymore with a corresponding internal resonance frequency of the scatterer. The superior accuracy of the desingularized boundary element method further ensures that the fictitious behavior is limited to very narrow bands in the frequency spectrum. The concepts were illustrated with examples of the scattering of a plane wave on a rigid sphere.

\section*{Acknowledgements}
This work is supported in part by the Australian Research Council grants, DE150100169 {\&} CE140100003, to QS and a Discovery Project Grant to DYCC.

\appendix
\section{Fictitious frequencies for the m=0 case} \label{sec:Appendix_A}

\begin{figure}[!t]
\centering\includegraphics[width=0.4\linewidth]{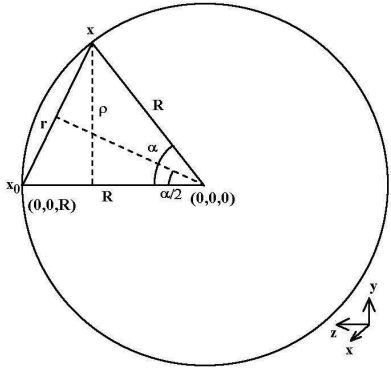}
\caption{Definition of the lengths $r=\| \boldsymbol x -\boldsymbol x_0 \|$, $\rho$, and the angles $\alpha$ and $\alpha/2$ for a sphere with radius $R$, it can easily be seen that $\sin(\alpha/2)=r/(2R)$ and $R\sin\alpha=\rho=r\cos(\alpha/2)$.}
\label{fig:Fig1}
\end{figure}
The normal derivative of the Green's function, $\partial G(\boldsymbol x ,\boldsymbol x_0|k)/
\partial n$, can be expressed as
\begin{equation} \label{eq:dGdn}
\frac{\partial G(\boldsymbol x ,\boldsymbol x_0|k)}
{\partial n} = (\boldsymbol x - \boldsymbol x_0) \cdot \boldsymbol n \frac{e^{ikr}}{r^3} (ikr-1).
\end{equation}
Without loss of generality lets assume that the point $\boldsymbol x_0$ is located on the z-axis (see also Fig. \ref{fig:Fig1} for the definition of symbols), thus $\boldsymbol x_0 = [0,0,R]$, the vectors $\boldsymbol x$ and $\boldsymbol n$ can then be presented by $\boldsymbol x=R[\cos\theta \sin\alpha, \sin\theta \sin\alpha, \cos\alpha]$ and
$\boldsymbol n=-\boldsymbol x /R$.
Then $(\boldsymbol x - \boldsymbol x_0) \cdot \boldsymbol n=R(-1+\cos\alpha)$. The surface element $\text{d}S= 2 \pi R  \rho \text{ d}\alpha$ can also be expressed as $\text{d}S= 2 \pi R^2 \sin \alpha\text{ d}\alpha$:
\begin{equation} \label{eq:BEM4}
\int_S
\frac{\partial G(\boldsymbol x ,\boldsymbol x_0|k)}
{\partial n} \text{ d}S = \int_0^{\pi}R[-1+\cos\alpha] \frac{e^{ikr}}{r^3}[ikr-1]2 \pi \sin\alpha   R^2 \text{ d}\alpha.
\end{equation}
With the help of Fig. \ref{fig:Fig1}, the term $(-1+\cos\alpha)$ can be rewritten as: $(-1+\cos\alpha)=-2\sin^2(\alpha/2)=-r^2/(2R^2)$. From $r=2R\sin(\alpha/2)$, one can deduce $R\text{ d}\alpha=\text{d}r/\cos(\alpha/2)$. With $\sin \alpha =\cos(\alpha/2)r/R$, the singular term $1/r^3$ in Eq. \ref{eq:BEM4} will be eliminated and this integral will turn into
\begin{equation} \label{eq:BEM5}
\int_S
\frac{\partial G(\boldsymbol x ,\boldsymbol x_0|k)}
{\partial n} \text{ d}S = -\frac{\pi}{R}\int_0^{2R}e^{ikr}[ikr-1]\text{ d}r.
\end{equation}
which will finally transform Eq. \ref{eq:BEM3} in:
\begin{equation} \label{eq:BEM6}
2\pi-2\pi\Big{\{}e^{i2kR}+\frac{1}{ikR}\big[-e^{i2kR}+1\big] \Big{\}}=0.
\end{equation}
Multiplying this equation by $e^{-ikr}$ and rearranging leads to
\begin{equation} \label{eq:Resonance1}
\sin(kR)[1-ikR]=0.
\end{equation}

\section{Fictitious frequencies for the m=1 case} \label{sec:Appendix_B}

In Section \ref{sec:m0} and \ref{sec:Appendix_A}, it was shown how the fictitious frequencies appear for the simplest case of $f=constant$, corresponding to the lowest order Legendre polynomials with $m=0$. The next least complicated function will be a linear function, corresponding to $m=1$. For simplicity sake, lets take $f=z$ as an example. Taking again $\boldsymbol x_0$ on the z-axis will give $f(\boldsymbol x_0)=R$ and with $c=2 \pi$, Eq. \ref{eq:BEM2} will turn into:
\begin{equation}
\label{eq:App1}
2\pi R+\int_S z
\frac{\partial G(\boldsymbol x ,\boldsymbol x_0|k)}
{\partial n} \text{ d}S = 0.    
\end{equation}
Eq. \ref{eq:BEM4} is still valid, except that an extra term $z=R\cos \alpha = R[1-r^2/(2R^2)]$ must be included, thus Eq. \ref{eq:BEM5} must be replaced by:
\begin{equation} \label{eq:App2}
\begin{aligned}
\int_S z
\frac{\partial G(\boldsymbol x ,\boldsymbol x_0|k)}
{\partial n} \text{ d}S = -\pi\int_0^{2R}\Big[1-\frac{r^2}{2R^2}\Big]e^{ikr}[ikr-1]\text{ d}r\\
=-\pi\int_0^{2R}e^{ikr}[ikr-1]\text{ d}r+\frac{\pi}{2R^2}\int_0^{2R}r^2e^{ikr}[ikr-1]\text{ d}r
\end{aligned}
\end{equation}
The first integral in the last expression is the same that appeared in Section \ref{sec:Appendix_A} as Eqs. \ref{eq:BEM5}, \ref{eq:BEM6} (except for a factor $1/R$), the second integral can be evaluated as:
\begin{equation} \label{eq:App3}
\begin{aligned}
\int_0^{2R}r^2e^{ikr}[ikr-1]\text{ d}r=8e^{i2kR} R^3\Big[1-\frac{2}{ikR}-\frac{2}{k^2R^2}+\frac{1}{ik^3R^3}\Big]-\frac{8}{ik^3}
\end{aligned}
\end{equation}
Thus Eq. \ref{eq:App2} becomes:
\begin{equation} \label{eq:App4}
\begin{aligned}
\int_S z
\frac{\partial G(\boldsymbol x ,\boldsymbol x_0)}
{\partial n} \text{ d}S = -2\pi R \Big{\{}e^{i2kR}+\frac{1}{ikR}\big[-e^{i2kR}+1\big] \Big{\}}\\+
4 \pi R e^{i2kR}\Big[1-\frac{2}{ikR}-\frac{2}{k^2R^2}+\frac{1}{ik^3R^3}\Big]-\frac{4 \pi R}{ik^3R^3}=-2 \pi R,
\end{aligned}
\end{equation}
where Eq. \ref{eq:App1} was used in the last equality. Multiplying by $e^{-ikR}/(2 \pi R)$ and regrouping terms with $e^{-ikR}$ and $e^{ikR}$ leads to:
\begin{equation} \label{eq:App5}
\begin{aligned}
e^{-ikR}\Big[1-\frac{1}{ikR}-\frac{2}{ik^3R^3}\Big]+e^{ikR}\Big[1-\frac{3}{ikR}-\frac{4}{k^2R^2}+\frac{2}{ik^3R^3}\Big]=0
\end{aligned}
\end{equation}
Expanding $e^{-ikR}$ and $e^{ikR}$ into $\cos(kR)$ and $\sin(kR)$ terms gives:
\begin{equation}\label{eq:App6}
\cos(kR)\Big[2-\frac{4}{ikR}-\frac{4}{k^2R^2}\Big] -i \sin(kR)\Big[\frac{2}{ikR}+\frac{4}{k^2R^2}-\frac{4}{ik^3R^3}\Big]   
\end{equation}
Separating this into real and imaginary parts: 
\begin{equation} \label{eq:App7}
\begin{aligned}
\text{Real part: }& \cos(kR)\Big[2-\frac{4}{k^2R^2}\Big]=\sin(kR)\Big[\frac{2}{kR}-\frac{4}{k^3R^3}\Big]
\\
\text{Imaginary part: }& \cos(kR)\frac{4}{kR}=\sin(kR)\frac{4}{k^2R^2}
\end{aligned}
\end{equation}
Both the real and imaginary part lead to the following condition:
\begin{equation}\label{eq:App8}
\tan(kR)=kR
\end{equation}
which is the same as the internal resonance condition for $m=1$, with solution $kR=4.49341$ etc. (see Table \ref{tab:Table1}).
\section{The m=0 case with a modified Green's function}\label{sec:Appendix_C}
The normal derivative of the additional part is: 
\begin{equation} \label{eq:BEM_Mod2}
\begin{aligned}
\frac{\partial G(\boldsymbol x, \boldsymbol a|k)}{\partial n} = \boldsymbol n \cdot [\boldsymbol x - \boldsymbol a] \frac{e^{ikr'}}{r'^3}[ikr'-1].
\end{aligned}
\end{equation}
As in Section \ref{sec:spuriousorigins}
assume that $f=const$ (corresponding to $m=0$) and again assume that the point $\boldsymbol x_0$ is located on the z-axis, the vectors $\boldsymbol x$ and $\boldsymbol n$ and $\text{d}S$ are defined the same as in Section \ref{sec:spuriousorigins}, while \\ $\boldsymbol x - \boldsymbol a =[R\cos\theta \sin\alpha, R\sin\theta \sin\alpha,R \cos\alpha -a]$. Thus $\boldsymbol n \cdot [\boldsymbol x - \boldsymbol a] = -R + a \cos \alpha$. For the length $r'$ the following relationship can be found:
\begin{equation} \label{eq:BEM_Mod3}
r'^2=R^2\sin^2\alpha +(R \cos\alpha -a)^2 =R^2-2aR\cos \alpha + a^2,    
\end{equation}
while for $\text{d}r'$ one finds:
\begin{equation} \label{eq:BEM_Mod4}
r'\text{ d}r'=aR\sin\alpha \text{ d}\alpha    
\end{equation}
Thus, similar to Eq. \ref{eq:BEM4}:
\begin{equation}
\label{eq:BEM_Mod5}
\int_S
\frac{\partial G(\boldsymbol x ,\boldsymbol a|k)}
{\partial n} \text{ d}S = \int_{R-a}^{R+a}\boldsymbol n \cdot [\boldsymbol x - \boldsymbol a] \frac{e^{ikr'}}{r'^3}[ikr'-1]2\pi r' \frac{R}{a}\text{ d}r'.
\end{equation}
Substituting $\boldsymbol n \cdot [\boldsymbol x - \boldsymbol a]= -R + a \cos \alpha$ and eliminating $\cos \alpha$ with Eq. \ref{eq:BEM_Mod3}:
\begin{equation}\label{eq:BEM_Mod6a}
\begin{aligned}
\int_S
\frac{\partial G(\boldsymbol x ,\boldsymbol a|k)}
{\partial n} \text{ d}S =& \frac{2\pi R}{a}\int_{R-a}^{R+a}
\Big[-R + \frac{R^2+a^2-r'^2}{2R} \Big]\frac{e^{ikr'}}{r'^2}[ikr'-1] \text{ d}r'\\
=&\frac{2\pi R}{a}\Big[1-\frac{1}{ikR}\Big]e^{ikR}[e^{ika}-e^{-ika}].
\end{aligned}
\end{equation}
The last equality can be obtained easiest by splitting the integral in two parts and using $\partial e^{ikr'}/\partial r' = e^{ikr'}[ikr'-1]/r'^2$. Eq. \ref{eq:BEM_Mod6a} can be simplified to:
\begin{equation}
\label{eq:BEM_Mod6}
\int_S
\frac{\partial G(\boldsymbol x ,\boldsymbol a|k)}
{\partial n} \text{ d}S = 
2\pi R\Big[1-\frac{1}{ikR}\Big]e^{ikR}2i \frac{\sin(ka)}{a}.
\end{equation}
Eq. \ref{eq:BEM6} will now have an additional part as:
\begin{equation} \label{eq:BEM_Mod7}
\begin{aligned}
2\pi-2\pi\Big{\{}e^{i2kR}+\frac{1}{ikR}\big[-e^{i2kR}+1\big] \Big{\}}\\
+c_2 2\pi R\Big[1+ \frac{1}{ikR} \Big]e^{ikR}2 i \frac{\sin(ka)}{a}=0.
\end{aligned}
\end{equation}
Multiplying by $e^{-ikR}/(4\pi i)$ gives:
\begin{equation} \label{eq:BEM_Mod8}
\begin{aligned}
\sin(kR)\Big[1+ \frac{1}{ikR} \Big]+
c_2 R \Big[1+ \frac{1}{ikR} \Big]\frac{\sin(ka)}{a}=0.
\end{aligned}
\end{equation}
Since the common term $[1+1/ikR]$ can never become zero (k is a real number), this finally simplifies to:
\begin{equation} \label{eq:BEM_Mod9b}
\sin(k R)+ c_2 (R/a) \sin(k a)=0.
\end{equation}
If this equation is satisfied for a certain wave number, $k$, then $k_f=k$.



\bibliographystyle{elsarticle-num-names}
\bibliography{references.bib}







\end{document}